\documentclass[sigplan,screen]{acmart}

\copyrightyear{2025}
\acmYear{2025}
\setcopyright{cc}
\setcctype{CC-BY}
\acmConference[ASPLOS '25]{Proceedings of the 30th ACM International Conference on Architectural Support for Programming Languages and Operating Systems, Volume 1}{March 30-April 3, 2025}{Rotterdam, Netherlands}
\acmBooktitle{Proceedings of the 30th ACM International Conference on Architectural Support for Programming Languages and Operating Systems, Volume 1 (ASPLOS '25), March 30-April 3, 2025, Rotterdam, Netherlands}\acmDOI{10.1145/3669940.3707216}
\acmISBN{979-8-4007-0698-1/25/03}


\usepackage[]{hyperref}


\usepackage{amsfonts}
\usepackage{amsmath}
\usepackage{amsthm}
\usepackage{cleveref}
\usepackage{graphicx}
\usepackage{xcolor}
\usepackage{listings}
\usepackage{subcaption}
\usepackage{tikz}
\usetikzlibrary{shapes.geometric, arrows, positioning, fit}
\usepackage{wrapfig}
\usepackage{pdfpages}
\usepackage[english]{babel}
\usepackage{multirow}
\addto\extrasenglish{
    
}
\addto\extrasenglish{
    
}
\usepackage{microtype}
\usepackage{balance}

\usepackage{makecell}

\theoremstyle{definition}
\newtheorem{theorem}{Theorem}
\newtheorem{definition}{Definition}

\definecolor{keywordcolor}{rgb}{0.5,0,0.5}
\definecolor{textgray}{gray}{0.4}
\definecolor{mygray}{rgb}{0.5,0.5,0.5}
\title{Composing Distributed Computations Through Task and Kernel Fusion}

\author{Rohan Yadav}
\affiliation{%
    \institution{Stanford University}
    \city{Stanford}
    \state{California}
    \country{USA}
}
\email{rohany@cs.stanford.edu}
\author{Shiv Sundram}
\affiliation{%
    \institution{Stanford University}
    \city{Stanford}
    \state{California}
    \country{USA}
}
\email{shiv1@stanford.edu}
\author{Wonchan Lee}
\affiliation{%
    \institution{NVIDIA}
    \city{Santa Clara}
    \state{California}
    \country{USA}
}
\email{wonchanl@nvidia.com}
\author{Michael Garland}
\affiliation{%
    \institution{NVIDIA}
    \city{Santa Clara}
    \state{California}
    \country{USA}
}
\email{mgarland@nvidia.com}
\author{Michael Bauer}
\affiliation{%
    \institution{NVIDIA}
    \city{Santa Clara}
    \state{California}
    \country{USA}
}
\email{mbauer@nvidia.com}
\author{Alex Aiken}
\affiliation{%
    \institution{Stanford University}
    \city{Stanford}
    \state{California}
    \country{USA}
}
\email{aiken@cs.stanford.edu}
\author{Fredrik Kjolstad}
\affiliation{%
    \institution{Stanford University}
    \city{Stanford}
    \state{California}
    \country{USA}
}
\email{kjolstad@cs.stanford.edu}

\renewcommand{\shortauthors}{Rohan Yadav et al.}

\newcommand{\name}{Diffuse}
\newcommand{\cunumeric}{cuPyNumeric}
\newcommand{\numpy}{NumPy}
\newcommand{\scipysparse}{SciPy Sparse}
\newcommand{\legatesparse}{Legate Sparse}

\definecolor{todocolor}{rgb}{0.8,0,0}
\definecolor{editcolor}{rgb}{0,0,0.8}
\newcommand{\TODO}[1]{{\color{todocolor}#1}}
\newcommand{\IGNORE}[1]{}

\newcommand{\bnfdef}{\mathrel{::=}}
\newcommand{\bnfalt}{\mathrel{\mid}}

\makeatletter
\gdef\@copyrightpermission{
  \begin{minipage}{0.3\columnwidth}
   \href{https://creativecommons.org/licenses/by/4.0/}{\includegraphics[width=0.90\textwidth]{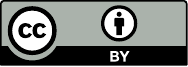}}
  \end{minipage}\hfill
  \begin{minipage}{0.7\columnwidth}
   \href{https://creativecommons.org/licenses/by/4.0/}{This work is licensed under a Creative Commons Attribution International 4.0 License.}
  \end{minipage}
  \vspace{5pt}
}
\makeatother

\begin{document}

\begin{abstract}

We introduce \name{}, a system that dynamically performs task and kernel 
fusion in distributed, task-based runtime systems.
The key component of \name{} is an intermediate representation of distributed computation 
that enables the necessary analyses for the fusion of distributed tasks to be performed 
in a scalable manner.
We pair task fusion with a JIT compiler to fuse together the kernels within fused tasks.
We show empirically that \name{}'s intermediate representation is general enough to be a target for
two real-world, task-based libraries (\cunumeric{} and Legate Sparse), letting
\name{} find optimization opportunities across function and library boundaries.
\name{} accelerates unmodified applications developed by composing task-based libraries
by 1.86x on average (geo-mean), and by between 0.93x--10.7x on up to 128 GPUs.
\name{} also finds optimization opportunities missed by the original application developers,
enabling high-level Python programs to match or exceed the performance of an explicitly
parallel MPI library.


\end{abstract}


\maketitle

\vspace{-0.65em}

\section{Introduction}

A modern trend in distributed programming is to develop drop-in implementations
of popular sequential libraries like \numpy{} or SciPy that automatically scale
to distributed machines while maintaining the semantics of the original library~\cite{legate, cunumeric, legate-sparse, numS}.
To achieve distribution, these drop-in replacement libraries are implemented by
translation to a distributed task-based runtime system~\cite{starpu, parsec, legion, pathways, ray}.
Libraries map computations to a stream of \emph{tasks} issued to the runtime, 
and map data on to runtime-managed distributed collections.
\emph{Tasks} are user-defined functions, whose bodies we call \emph{kernels},
that operate on subsets of the distributed collections.
The runtime is responsible for extracting parallelism from 
the input sequence of tasks and for computing the synchronization
and communication required between tasks.
This architecture enables distributed libraries to be built independently
and then composed freely, as the runtime system is responsible
for scheduling parallel work and maintaining coherence of distributed data.

However, the same abstractions that yield important composition properties
internally and externally to these distributed libraries can result in
degraded end-to-end performance.
The task decomposition of library operations results in
tasks that may be optimized individually but can have poor data locality and allocate much more temporary data than a different program organization that breaks down the abstraction boundaries by fusing tasks together both
within the operations of a particular library and across library boundaries.
As the task-based runtime system is issued a stream of tasks after library
abstraction boundaries have been traversed, the runtime has the opportunity
to fuse the tasks from different libraries together, which in turn enables
the fusion of the kernels of nested loops within fused tasks.
Fusion by the task-based runtime allows for these optimizations
to be performed without being limited to the semantics of any particular task-based library.

%
Prior works such as Weld~\cite{weld} and Split Annotations~\cite{split-annotations}
have developed techniques to perform fusion across library boundaries, but only
for shared memory libraries.
Distributed memory complicates program analyses, as
distributed data requires
communication when shared data is written to and read from by different nodes.
%
For example, a sequence of element-wise operations on a pair of 
distributed arrays may or may not be fusible depending on whether
the arrays are aliases of the same distributed data.
We do not consider the problem of automatic
parallelization~\cite{adve-crummey-par, uday-thesis, saman-par, wolf-par, wonchan-par}; the task-based programs we consider are already (implicitly) parallel.
We focus on the efficient composition of independently-written parallel, distributed programs.

We present \name{}, a system that dynamically performs task and kernel fusion for distributed, task-based
runtime systems, transparently achieving optimizations found in hand-tuned programs.
%
%
\name{} reasons over a task-based IR of distributed
computation, modeling computation as a sequence of tasks operating
on partitioned data (\Cref{sec:ir}).
\name{}'s IR is \emph{scale-free}, meaning that the size of the
IR and analyses on it are independent of the size of the target machine.
\name{} uses this IR to perform a dynamic dependence analysis to
fuse tasks in a distributed-memory setting (\Cref{sec:dist-task-fusion}).
\name{} pairs task fusion with a JIT compiler based on MLIR~\cite{mlir} that fuses and optimizes 
kernels within dynamically fused tasks, enabling data reuse across independent tasks (\Cref{sec:kernel-fusion}).
By analyzing a task-based IR, \name{}'s optimizations are not tied to the semantics
of any particular library.


We implement \name{} as a middle layer between high-level task-based libraries
and the low-level Legion runtime system~\cite{legion}.
To demonstrate \name{}, we modify the implementations of the distributed libraries \cunumeric{}~\cite{cunumeric} and Legate Sparse~\cite{legate-sparse} 
to target \name{}'s IR, and to expose 
their task implementations in MLIR for \name{}'s compiler to process.
\name{} then performs dynamic analyses to fuse the tasks and 
kernels issued by these libraries before forwarding the optimized
tasks to Legion.
As a result, programmers using \cunumeric{} and Legate Sparse benefit from \name{}
without modifying their applications.

To evaluate \name{}, we apply it to micro-benchmarks and several full scientific 
computing applications developed in \cunumeric{} and Legate Sparse,
including sparse Krylov solvers and physical simulations.
We compare against the standard implementations of \cunumeric{} and Legate Sparse and show that
\name{} achieves 1.86x speedup on average (geo-mean) over unmodified applications on up to 128 GPUs.
We additionally compare against the high-performance MPI-based PETSc~\cite{petsc} library and 
show that  \name{} enables naturally written \numpy{} and \scipysparse{} programs to match or 
exceed the performance of PETSc (1.4x geo-mean speedup).
Finally, we show that \name{} is able to find fusion and optimization opportunities missed by
the original application developers, achieving 1.23x 
speedup on average (geo-mean) over already hand-optimized code.

\section{Motivating Example}\label{sec:motivating-example}

\begin{figure}
\begin{subfigure}[h]{\linewidth}

\begin{subfigure}[h]{0.49\textwidth}
\begin{center}
\begin{tabular}{c}
\begin{lstlisting}
import cunumeric as np
grid = np.random.rand((N+2,N+2))
# Create multiple aliasing views
# of the distributed grid array.
center = grid[1:-1, 1:-1]
north  = grid[0:-2, 1:-1]
east   = grid[1:-1, 2:  ]
west   = grid[1:-1, 0:-2]
south  = grid[2:  , 1:-1]
for i in range(niters):(*\label{line:loop-start}*)
  avg = center + north + \(*\label{line:avg-computation}*)
        east + west + south
  work = 0.2 * avg(*\label{line:avg-computation-end}*)
  center[:] = work(*\label{line:loop-end}*)
\end{lstlisting}
\end{tabular}
\end{center}
\caption{\cunumeric{} source code.}
\label{fig:cunumeric-stencil}
\end{subfigure}
\begin{subfigure}[h]{0.49\textwidth}
    \centering
    \includegraphics[width=0.9\textwidth]{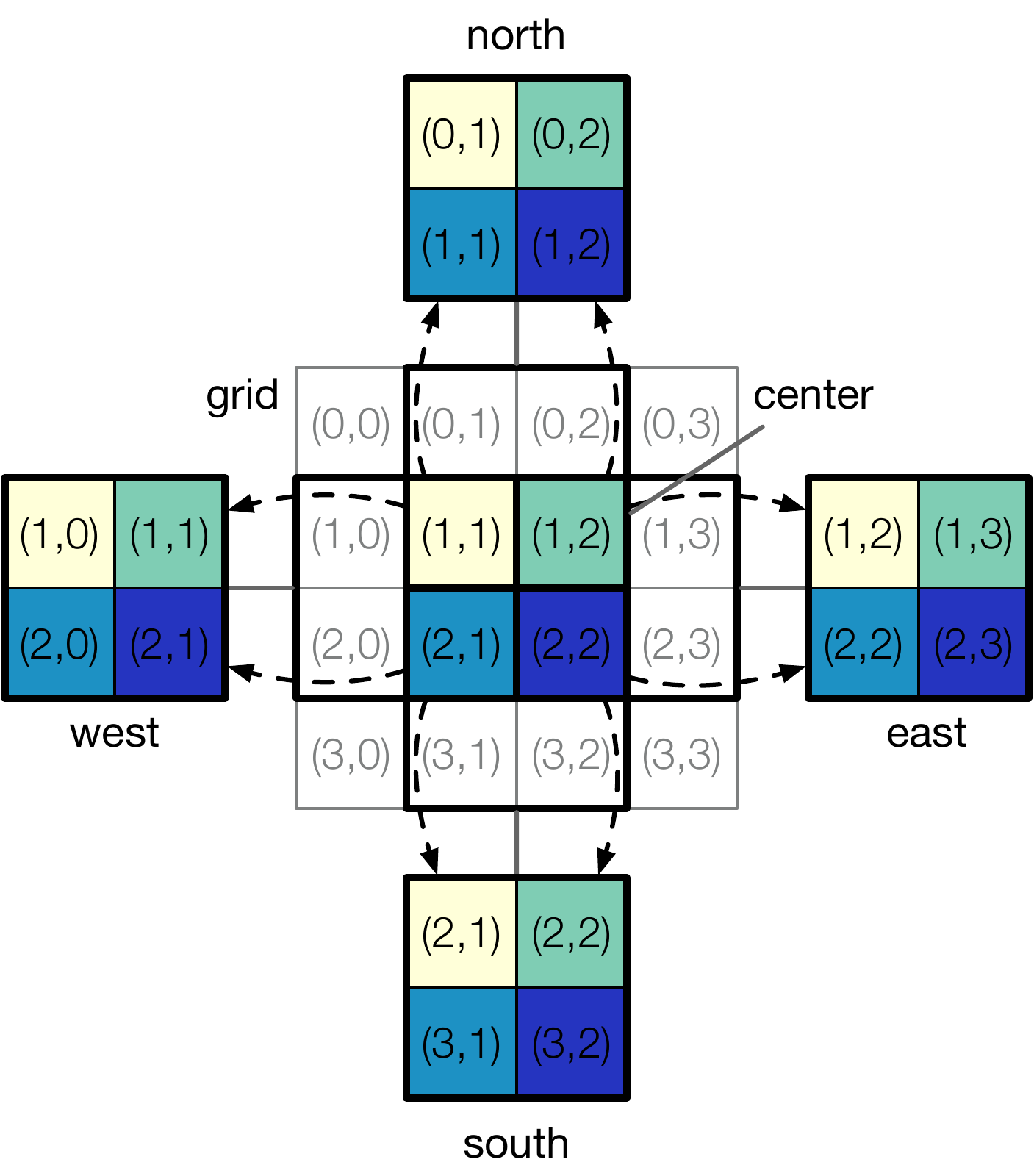}
    \caption{4-node execution, colors denote cells held by each node.
    Dotted lines denote communication.}
    \label{fig:stencil-visualization}
\end{subfigure}

\begin{subfigure}[h]{0.47\textwidth}
\begin{lstlisting}
# ADD, MULT and COPY are in
# cuNumeric's implementation.
ALLOC ARRAY t1
ADD(center, north, t1)
ALLOC ARRAY t2
ADD(t1, east, t2)
ALLOC ARRAY t3
ADD(t2, west, t3)
ALLOC ARRAY avg
ADD(t3, south, avg)
ALLOC ARRAY work
MULT(0.2, avg, work)
COPY(work, center)
\end{lstlisting}
\caption{Stream of tasks and allocations issued by the main loop.}
\label{fig:cunumeric-stencil-task-stream}
\end{subfigure}
\hfill
\begin{subfigure}[h]{0.47\textwidth}
\begin{lstlisting}
# FUSED_ADD_MULT is a new
# task generated by Diffuse.
ALLOC ARRAY work
FUSED_ADD_MULT(
  center, 
  north, 
  east, 
  west, 
  south, 
  0.2, 
  work
)
COPY(work, center)
\end{lstlisting} 
\caption{Operation stream after \name{}'s optimization.}
\label{fig:cunumeric-stencil-optimized-stream}
\end{subfigure}

\begin{subfigure}[h]{0.47\textwidth}
\begin{lstlisting}
# ADD, MULT and COPY are
# elementwise operators.
def ADD(a, b, c):
  for i, j in a:
    c[i,j] = a[i,j] + b[i,j]
def MULT(s, a, b):
  for i, j in a:
    b[i,j] = s * a[i,j]
def COPY(a, b):
  for i, j in a:
    b[i,j] = a[i,j]
\end{lstlisting}
\caption{Tasks invoked during standard execution.}
\label{fig:cunumeric-stencil-tasks}
\end{subfigure}
\hfill
\begin{subfigure}[h]{0.47\textwidth}
\begin{lstlisting}
# FUSED_ADD_MULT performs
# the scaled five-way add.
def FUSED_ADD_MULT(
  a, b, c, d, e, s, out
):
  for i, j in a:
    out[i,j] = s * (
        a[i,j] + b[i,j] 
      + c[i,j] + d[i,j]
      + e[i,j])
\end{lstlisting}
\caption{Fused task generated by \name{}.}
\label{fig:cunumeric-stencil-optimized-tasks}
\end{subfigure}

\end{subfigure}

\caption{Execution example of \name{} on a distributed, multi-GPU \cunumeric{} 5-point stencil application.}
\label{fig:intro-cunumeric-stencil-example}

\end{figure}

\Cref{fig:intro-cunumeric-stencil-example} shows how \name{} optimizes the
 \cunumeric{} program in \Cref{fig:cunumeric-stencil} that performs
a 5-point stencil computation.
The \cunumeric{} library is a drop-in replacement for \numpy{}~\cite{numpy} that scales unmodified \numpy{}
programs to distributed machines by targeting the Legion~\cite{legion} runtime system.
\cunumeric{} maps \numpy{} arrays to Legion's regions, and maps 
\numpy{} functions to task launches operating on regions that
are partitioned across the machine.
As the \cunumeric{} program executes, it issues
a stream of tasks to the Legion runtime, which dynamically
discovers the necessary communication and synchronization 
required to execute the tasks on the target machine.
The program execution on a four-by-four grid with four nodes is visualized in \Cref{fig:stencil-visualization},
where each node owns an element of each aliasing view of the \texttt{grid} array.
The dotted arrows represent the communication required to propagate updates to the \texttt{center} array to the other aliasing
views of \texttt{grid}.
\Cref{fig:cunumeric-stencil-task-stream} is a simplified representation of the task stream
that \cunumeric{} issues during execution of the inner loop (lines \ref{line:loop-start}--\ref{line:loop-end} of \Cref{fig:cunumeric-stencil}), and \Cref{fig:cunumeric-stencil-tasks} contains
pseudocode for each of the task implementations.
This stream of operations creates multiple temporary distributed arrays 
for the results of individual operations, and separate tasks for each corresponding 
addition and multiplication.
The combination of temporary arrays and separate tasks of loops is an
inefficient execution strategy.
\name{} speeds this program up by four times by
creating a new fused task that
computes the \texttt{work} array 
(lines \ref{line:avg-computation}-\ref{line:avg-computation-end}) in a single operation and removes
the temporary arrays, including \texttt{avg}, resulting in the stream of operations
in \Cref{fig:cunumeric-stencil-optimized-stream} and the generated fused task in \Cref{fig:cunumeric-stencil-optimized-tasks}.
Interestingly, \name{} does not fuse the task that performs
\texttt{center[:] = work} (line \ref{line:loop-end} of \Cref{fig:cunumeric-stencil}).

To understand these decisions, we must introduce the distributed aspect
of the tasks and data collections in \Cref{fig:cunumeric-stencil-task-stream}.
Each task in \Cref{fig:cunumeric-stencil-task-stream} actually represents a group of parallel
tasks launched over partitioned arrays, where each parallel task operates 
on a subset of the partitioned data.
Dependencies and communication that arise from parallel tasks operating
on the same distributed data affect when fusion is possible.
In our example, the arrays \texttt{center}, \texttt{north}, \texttt{east}, 
\texttt{west}, and \texttt{south} are aliasing views of the array \texttt{grid},
meaning that they share logical array entries.
Because these distributed arrays alias, \name{} does not fuse the task group that computes
\texttt{center[:] = work} into the task group that reads from \texttt{north}, \texttt{east},
\texttt{west} and \texttt{south}, as the fusion would create a task group that concurrently
reads and writes to aliasing data.
Similarly, the \texttt{center[:] = work} task group issued at iteration $i$ cannot
be fused into the \texttt{avg} computation (line \ref{line:avg-computation} of \Cref{fig:cunumeric-stencil})
at iteration $i+1$ because communication is required to propagate updates to \texttt{center}.
To reason about distributed computations over partitioned data, we develop a
scale-free intermediate representation (\Cref{sec:ir}) that models tasking
runtime systems which support aliased views of distributed data.
We then develop a dynamic analysis for task 
fusion (\Cref{sec:dist-task-fusion}) that
reasons about dynamically known communication patterns 
in distributed computations to fuse groups of parallel tasks.

\section{Intermediate Representation}\label{sec:ir}

\begin{figure}

\begin{subfigure}[h]{1.0\linewidth}
    \footnotesize
    \centering{\textbf{Syntax}}
    \[
        \begin{array}{rlcl}
        \textit{Unique ID} & id & & \\
        \textit{Point} & p & \bnfdef & (\mathbb{Z}, \ldots) \\
        \\
        
        \textit{Store} & S & \bnfdef & \textsf{Store}(id, p) \\
        \textit{Projection Function} & F & \bnfdef & \textsf{Projection}(id, \textsf{Point} \rightarrow \textsf{Point}) \\
        \textit{Partition} & P & \bnfdef & \textsf{None} \bnfalt \textsf{Tiling}(p, p, F) \\
        \\

            \textit{Privilege} & Pr & \bnfdef & \textsf{Read (R)} \bnfalt \textsf{Write (W)} \bnfalt\\
                                          &&& \textsf{Reduce (Rd)} \bnfalt \textsf{Read-Write (RW)}\\
        \textit{Index Task} & T & \bnfdef & \textsf{IndexTask}(p, (S, P, Pr)~\textsf{list}) \\
        \textit{Task Window} & W & \bnfdef & T~\textsf{stream} \\
        \end{array}
    \]

    \quad
    
    \textbf{Constructs for Reasoning}
    \[
        \begin{array}{rlcl}
        \textit{Sub-Store} & S^p & \triangleq & \textsf{SubStore}(S, P, p) \\
        \textit{Point Task} & T^p & \triangleq & \textsf{PointTask}((S^p, Pr)~\textsf{list}) \\
        \end{array}
    \]
    \caption{\name{}'s intermediate representation.}
    \label{fig:ir-grammar}
\end{subfigure}

\begin{subfigure}[h]{1.0\linewidth}
\centering
\includegraphics[width=0.9\textwidth]{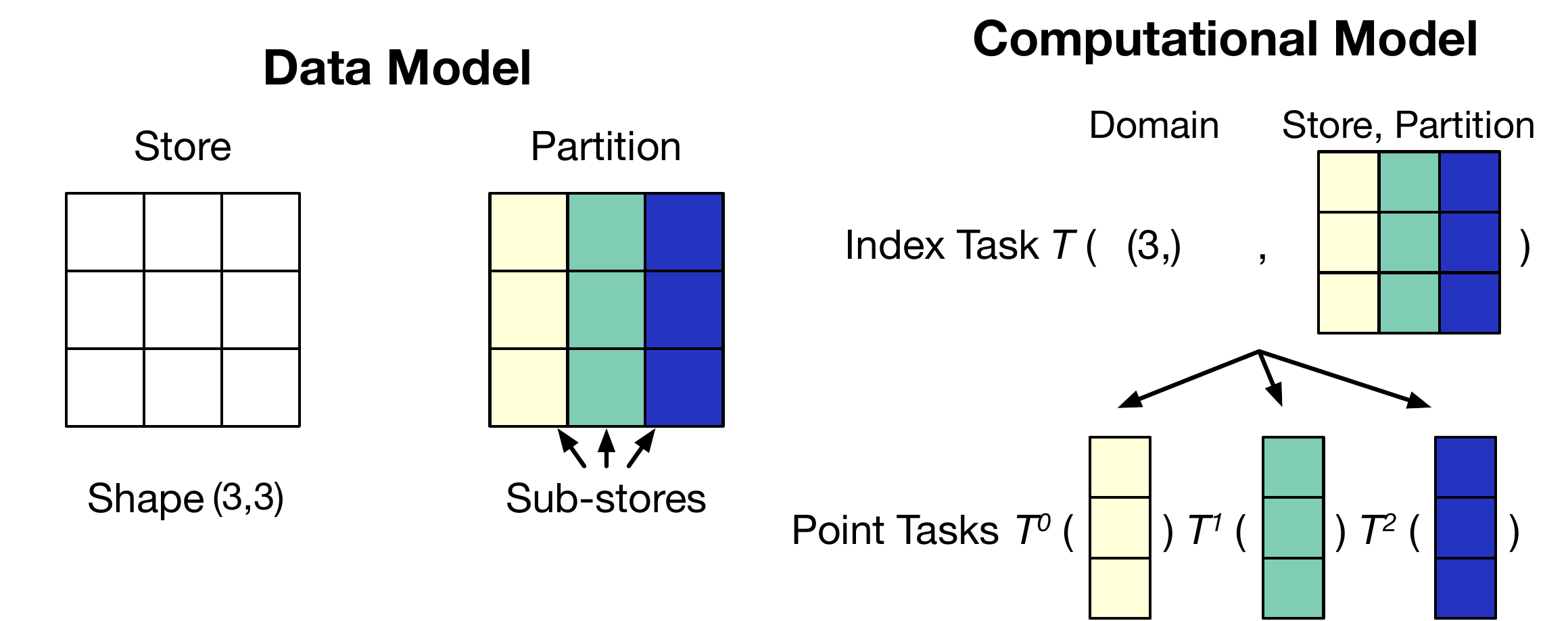}
\caption{Relationships between components of \name{}'s IR.}
\label{fig:ir-viz}
\end{subfigure}
\caption{\name{}'s IR exposes a distributed data model and a model for distributed computation on distributed data.}
\label{fig:ir}
\end{figure}

The first contribution of \name{} is an IR that enables scalable fusion analyses
through a \emph{scale-free}
representation of distributed programs, meaning that the size of the representation
is independent of the total number of processors in the target system.
\name{}'s IR is an abstraction over the collections of concrete tasks
and distributed data structures of a lower-level task-based programming system, like Legion,
that usually have \emph{scale-aware} representations.
We have modified \cunumeric{} and \legatesparse{} to dynamically
generate programs in \name{}'s IR instead of targeting
Legion directly.
%
\name{}'s IR, presented in \Cref{fig:ir}, is designed
to make it inexpensive to perform the analyses required for fusion, while still being able to express sophisticated computations.
The IR contains a data model to represent distributed data, and a computational
model to define distributed computations over distributed data.
The syntax of the IR is in \Cref{fig:ir-grammar}, and a visualization of the
IR's structure is shown in \Cref{fig:ir-viz}.

\subsection{Data Model}

\name{} represents distributed data as \emph{stores}, which are distributed arrays.
Each store has a unique ID and a rectangular shape defined by a tuple of non-negative integers,
representing the upper bound of each dimension of the store.
We refer to these rectangular shapes as \emph{domains}, which are also used to describe the decomposition
of data and compute across processors.
Stores are partitioned across the target machine into \emph{sub-stores}, which are subsets
of a store.

Partitions of stores are first-class objects in \name{}.
A \emph{partition} is a mapping from points in a domain to sub-stores, where each
point in the domain represents a processor.
This mapping is represented by \name{} in a structured manner, breaking different kinds
of mappings into different syntactic groups.
For simplicity of presentation, we consider two kinds of partitions, sufficient
to explore the analyses used in \name{}.
%
Our implementation supports more partition kinds with no
additional technical insights.
The main requirement on partitions is that two partitions of the same kind can be checked for inequality without examining each sub-store within each partition.
This requirement is critical for a scalable analysis,
as discussed in \Cref{sec:dist-task-fusion}.

\begin{figure}
\begin{subfigure}[t]{0.23\linewidth}
\centering
\includegraphics[width=\textwidth]{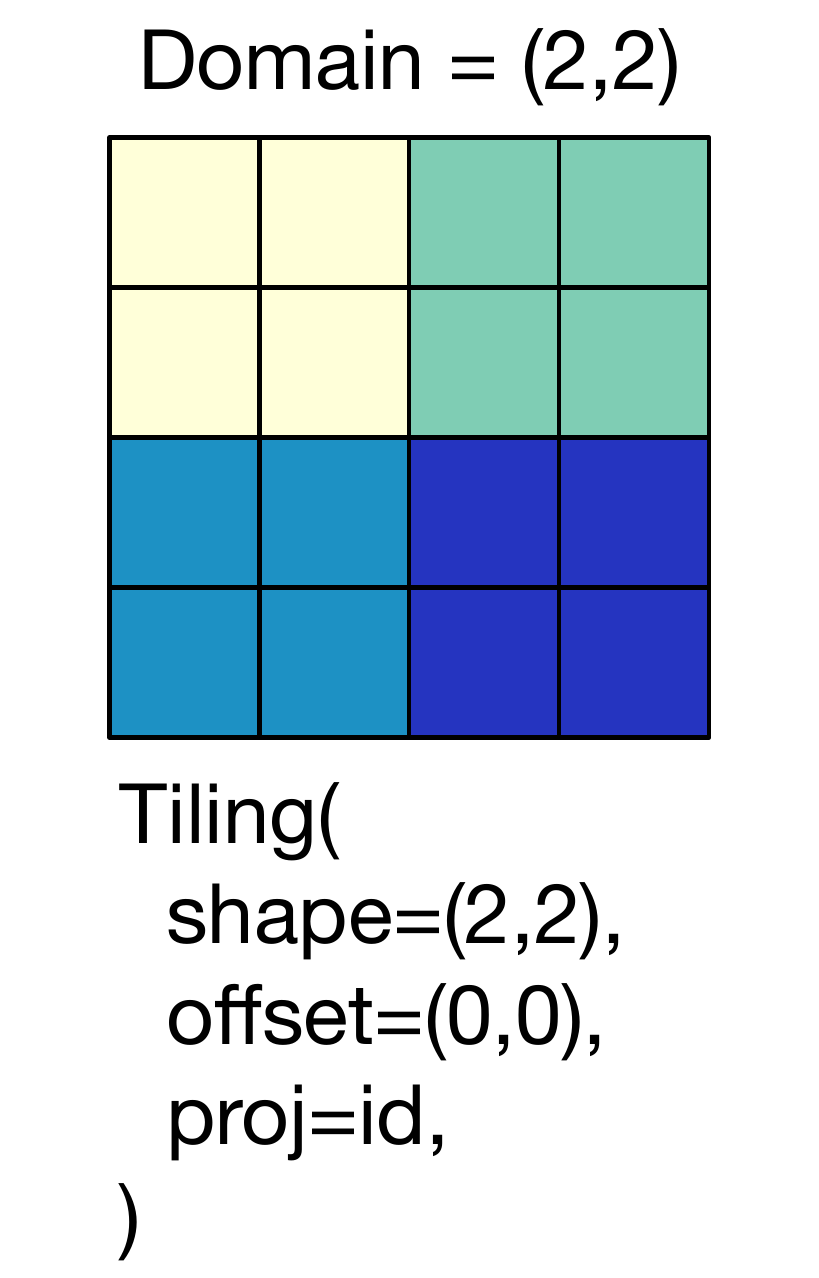}
\caption{2x2 tiling of a 4x4 store.}
\label{fig:tiled-part-1}
\end{subfigure}
\hfill
\begin{subfigure}[t]{0.23\linewidth}
\centering
\includegraphics[width=\textwidth]{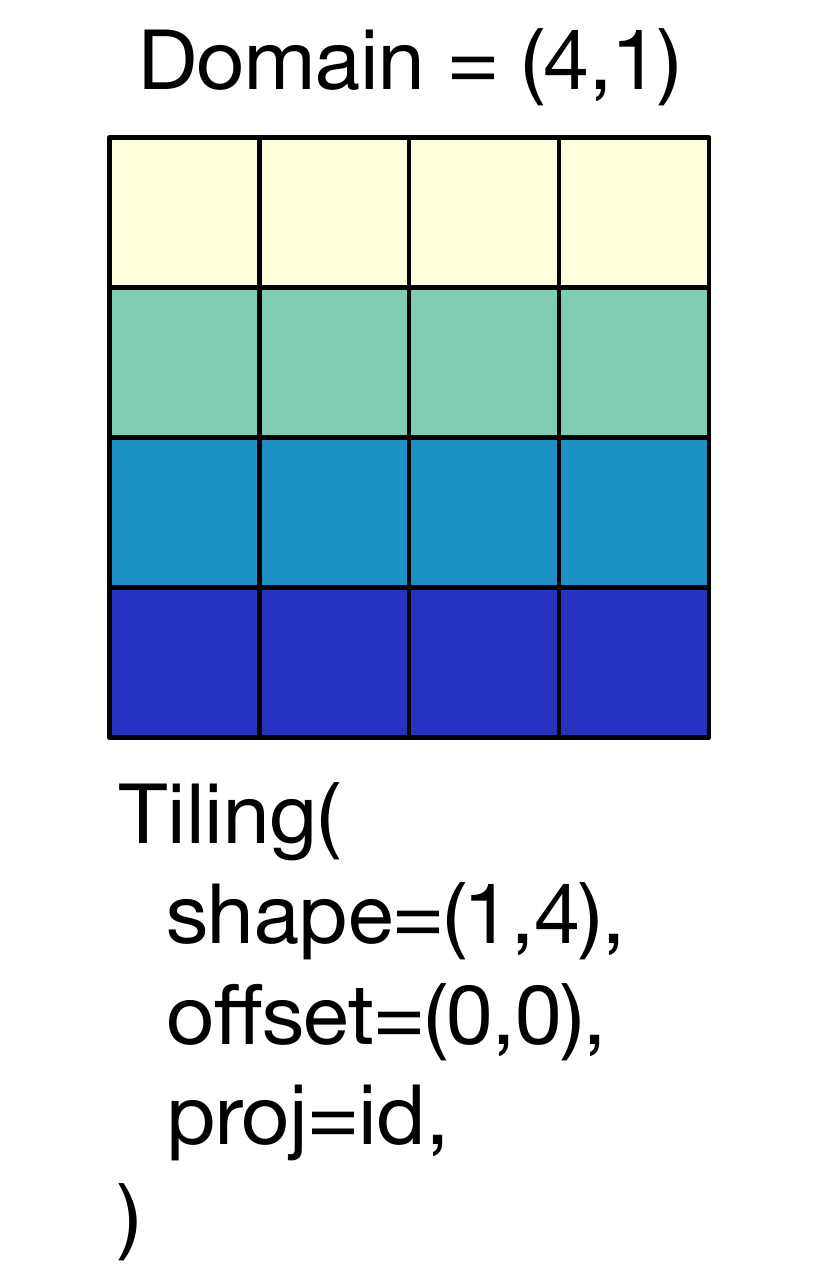}
\caption{1x4 tiling of a 4x4 store.}
\label{fig:tiled-part-2}
\end{subfigure}
\hfill
\begin{subfigure}[t]{0.23\linewidth}
\centering
\includegraphics[width=\textwidth]{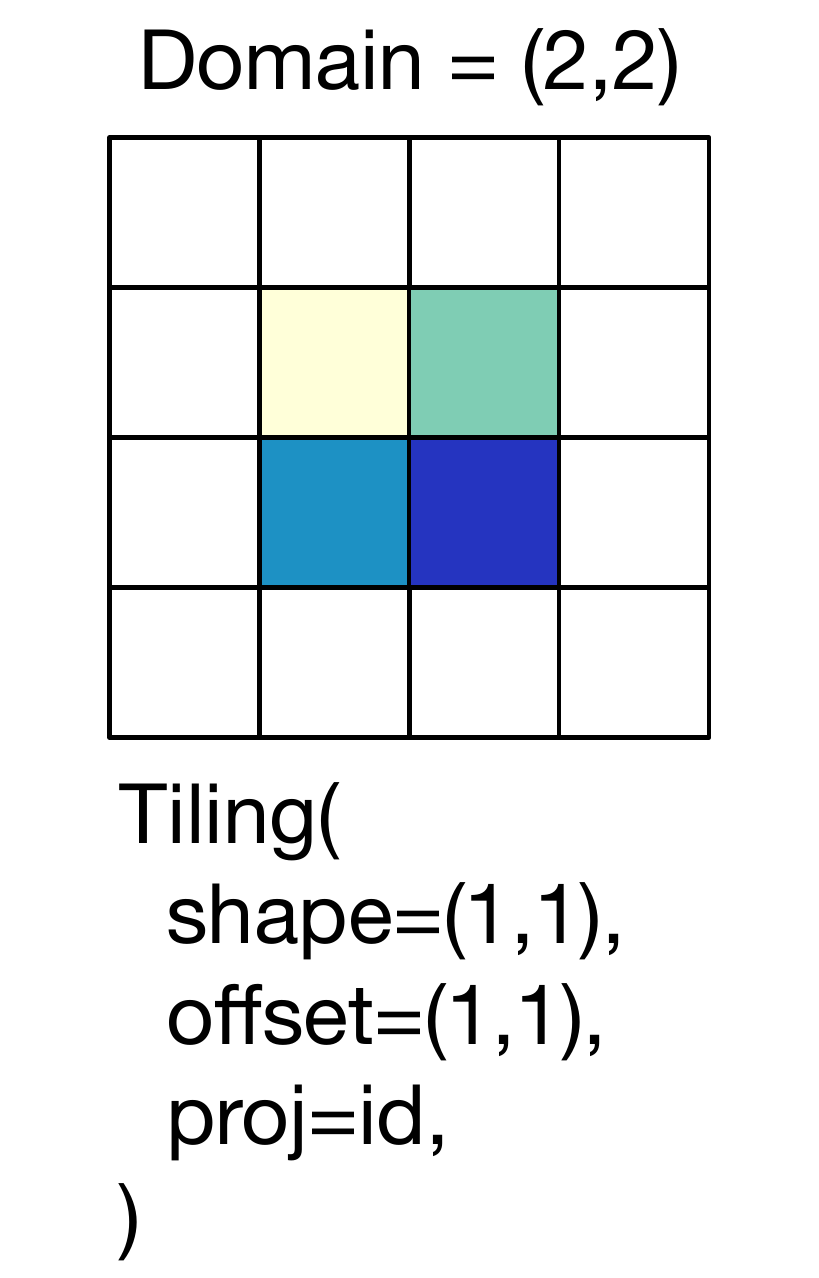}
\caption{Offset 1x1 tiling of a 4x4 store.}
\label{fig:tiled-part-3}
\end{subfigure}
\hfill
\begin{subfigure}[t]{0.23\linewidth}
\centering
\includegraphics[width=\textwidth]{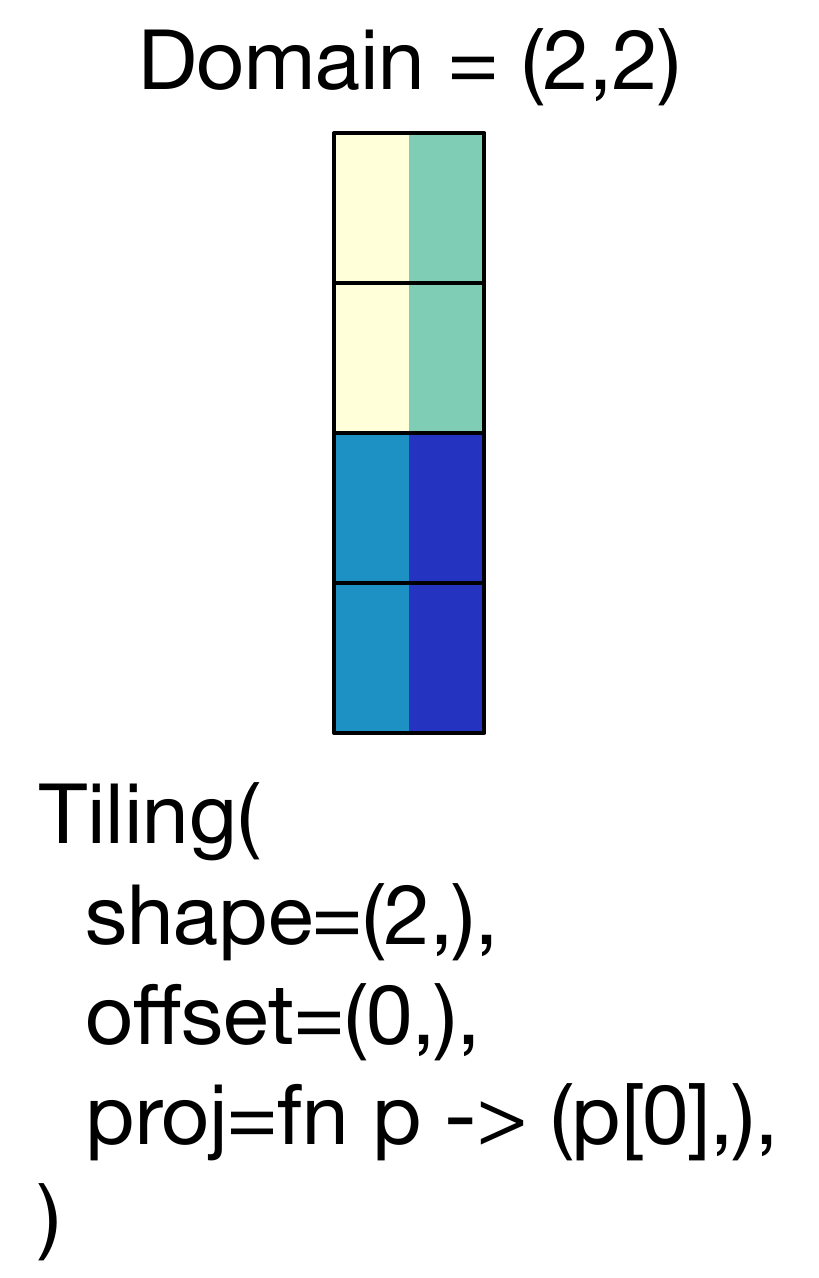}
\caption{Aliased blocking of a size 4 store.}
\label{fig:tiled-part-4}
\end{subfigure}

\quad
\begin{subfigure}[h]{0.9\linewidth}
\centering
\footnotesize
\[
\textsf{sub-store-bounds}(\textsf{Tiling}(\textsf{shape}, \textsf{offset}, \textsf{proj}), p) =
\]
\[
[\textsf{proj}(p) * \textsf{shape}, \textsf{proj}(p + 1) * \textsf{shape}) + \textsf{offset}
\]
\hfill
\caption{Function that computes a bounding-box within the store that a \textsf{Tiling} partition maps point $p$ to.}
\label{fig:tiling-point-func}
\end{subfigure}

\caption{Examples of \textsf{Tiling} partitions. Partitions maps points in the denoted domain to sub-stores. Each color refers to the sub-store associated with a each point in the domain.}
\label{fig:tiling-examples}
\end{figure}

The first partition kind \textsf{None} represents the replication of a store,
where all points in the partition's domain are mapped to the entire store.
The second partition kind \textsf{Tiling} represents an $n$-dimensional affine tiling of a store.
A \textsf{Tiling} contains an $n$-dimensional tile shape and an offset from the origin,
which are used to compute the sub-store associated with each point
in the partition's domain.
For example, \Cref{fig:tiled-part-1} shows a tiling of a two-dimensional store using 2x2 tiles
over a 2x2 domain,
while \Cref{fig:tiled-part-2} shows a row-based tiling (i.e., tiles of size 1x4) of the same store over a 4x1 domain.
\Cref{fig:tiled-part-3} shows a partition of a subset of the store beginning at the point $(1, 1)$.
\textsf{Tiling} partitions also contain a \emph{projection function} that applies a transformation
to each point in the partition's domain before computing the subset with the tile size and offset.
Projection functions enable \textsf{Tiling} partitions to express aliased and replicated data.
For example, \Cref{fig:tiled-part-4} shows a vector tiled over a two-dimensional domain by a
 projection function that discards the second dimension of each point in the partition's domain, resulting in
a partially aliased partition.
The formula that defines the sub-store bounds for each point of a \textsf{Tiling} partition
is shown in \Cref{fig:tiling-point-func}.
The representations of \textsf{None} and \textsf{Tiling} partitions are scale-free
as the mapping of points to sub-stores is implicit in the partition, rather than
explicitly storing the bounds of each sub-store in the partition.

%
%

To reason about the sub-store referenced by each point of
a partition, we include an explicit $\textsf{SubStore}(S, P, p)$ construct, 
representing the sub-store associated with point $p$ of store $S$ using partition $P$.
As a short-hand, we let $S[P, p] = \textsf{SubStore}(S, P, p)$, and
refer to $S$ as the \emph{parent} store of $S[P, p]$.
The indices contained within the sub-store $S[P, p]$ are directly computable in cases
when $P$ is \textsf{None} or \textsf{Tiling}, but may depend on runtime values
held by stores when more complex partitioning operators are introduced.
%
%
Our later definitions assume that it is possible to find the intersection between
two sub-stores, but our fusion algorithm in \Cref{sec:dist-task-fusion} does not require explicit
computation of these intersections.

\subsection{Computational Model}

%
\name{} models computation as a stream of \emph{index tasks}~\cite{index-launches}
issued in program order.
%
An $\textsf{IndexTask}(d, A)$ represents a group of parallel 
tasks over points in a rectangular domain $d$, referred to as the \emph{launch domain}.
An index task operates on the list $A$ of stores, partitions, and privileges,
using the denoted privilege to access the requested partition of each store.
We refer to each privilege with the abbreviations noted in parentheses.
%
Each parallel task within the group reads from, writes to, 
or reduces to the sub-stores referred to by the stores and partitions at each point.
The parallel tasks within an index task group may perform arbitrary
computation on argument stores that respects the requested
privilege on each argument store.
For the simplicity of presentation, we assume that the \textsf{Reduce} privilege 
refers to a single reduction function being applied (such as addition).
%
%
%
This representation is explicitly parallel as tasks are annotated with their launch domain and 
partitions of distributed data structures.
However, the representation is scale-free as the size of the representation
is independent of the degree of parallelism --- only the symbolic
size of the launch domain increases.

Similar to sub-stores, \name{}'s IR
has a notion of a \emph{point task}, which is one point in an index task's launch domain.
Given an index task $T = \textsf{IndexTask}(d, A)$, let $T^p$ be the point task at point $p \in d$, 
operating on the list of stores $[(S[P, p], pr) : \forall~(S, P, pr) \in A]$.
Point tasks operate on the sub-stores corresponding to their index point.

We define the predicates $\textsf{R}(T, (S, P))$, $\textsf{W}(T, (S, P))$ and $\textsf{Rd}(T, (S, P))$
to be true when the task $T$ correspondingly reads from, writes to, or reduces to the store $S$ using partition $P$.
When $(S, P)$ has the privilege \textsf{Read-Write}, both $\textsf{R}(T, (S, P))$ and $\textsf{W}(T, (S, P))$ are true.
We also overload these predicates for point tasks and sub-stores, where $\textsf{R}(T^p, S)$ is true
when point task $T^p$ reads sub-store $S$.

The dynamic semantics of \name{}'s IR are defined by a translation to an underlying task-based
runtime system such as Legion~\cite{legion}.
Stores are mapped to the distributed data structures of the underlying runtime system,
and \name{}'s first-class, structured partitions are mapped onto lower-level, unstructured partitions.
Finally, index tasks are translated to tasks in the lower-level runtime
system and issued for execution.

\section{Distributed Task Fusion}\label{sec:dist-task-fusion}

\name{} leverages this IR to fuse distributed computations
through task fusion, enabling the fusion of kernels 
within fused tasks (\Cref{sec:kernel-fusion}).
%
Applications submit index tasks to \name{}, which
buffers the tasks into a \emph{window} of tasks
to be analyzed before submission to the underlying runtime.
A distributed task fusion algorithm finds a fusible prefix of
tasks in the window, and replaces the prefix with a fused task.
To be fusible, the prefix of index tasks must
be executable in sequence without cross-processor communication.
%
%
We define when communication may occur between index tasks and describe when 
a sequence of index tasks is fusible.
We then give an algorithm for finding fusible index task sequences.

\subsection{Dependencies}\label{sec:dependencies}

Dependencies are well-studied---we discuss how to
define dependencies between \name{}'s index tasks.
We adopt the terminology of Aho et al.~\cite{dragon-book} when possible.
Communication is required between point tasks that have a dependence.
The dependence implies synchronization and potentially data movement between the point tasks.
A dependency exists between two point tasks that access the same data unless
both tasks read from or reduce to the data with the same associate and commutative operator.
Recall that for an index task $T$, we refer to the point task at point $p$ as $T^p$.
We define $\textsf{dep}(T_1^p, T_2^{p'})$ to be true if $T_2^{p'}$ depends on $T_1^p$.
%
%
\begin{definition}\label{def:dependencies}
    Given point tasks $T_1^{p}$, $T_2^{p'}$ where index task $T_1$ is issued before index task $T_2$,
    $\textsf{dep}(T_1^{p}, T_2^{p'})$ if 
    $~\exists$ sub-stores $S, S'$ 
    with the same parent such that 
    $S \cap S' \neq \emptyset$ and either
    \begin{description}
        \item[true-dep:] \hfill
        
        $\textsf{W}(T_1^{p}, S) \wedge
        \left(\textsf{R}(T_2^{p'}, S') \vee \textsf{W}(T_2^{p'}, S') \vee \textsf{Rd}(T_2^{p'}, S')\right)$
        \item[anti-dep:] \hfill
        
        $\textsf{R}(T_1^{p}, S) \wedge \left(\textsf{W}(T_2^{p'}, S') \vee \textsf{Rd}(T_2^{p'}, S')\right)$
        \item[reduction-dep:] \hfill
        
        $\textsf{Rd}(T_1^{p}, S) \wedge \left(\textsf{R}(T_2^{p'}, S') \vee \textsf{W}(T_2^{p'}, S')\right)$.
    \end{description}
\end{definition}

\begin{figure}
\hfill
\begin{subfigure}[h]{\linewidth}
\begin{subfigure}[h]{0.48\textwidth}
    \centering\captionsetup{justification=centering}
    \includegraphics[width=0.75\textwidth]{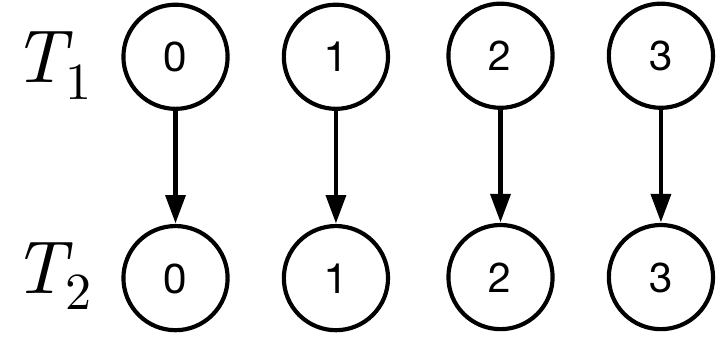}
    \caption{Point-wise dependence map: $\mathcal{D}(T_1, T_2)[p] = \{p\}$}
\end{subfigure}
\begin{subfigure}[h]{0.48\textwidth}
    \centering\captionsetup{justification=centering}
    \includegraphics[width=0.75\textwidth]{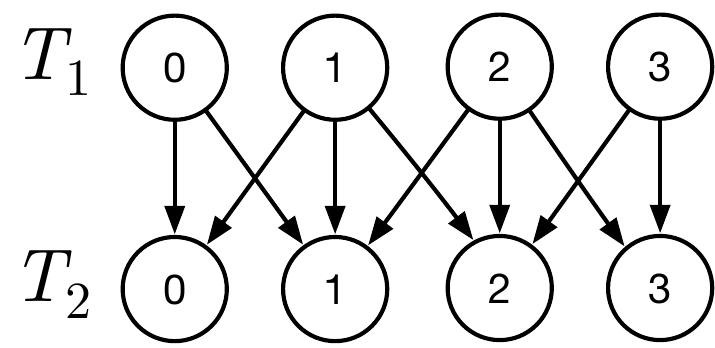}
    \caption{Stencil dependence map: $\mathcal{D}(T_1, T_2)[p] = \{p-1, p, p+1\}$}
\end{subfigure}
\end{subfigure}

\begin{subfigure}[h]{\linewidth}
    \centering
    \includegraphics[width=0.7\textwidth]{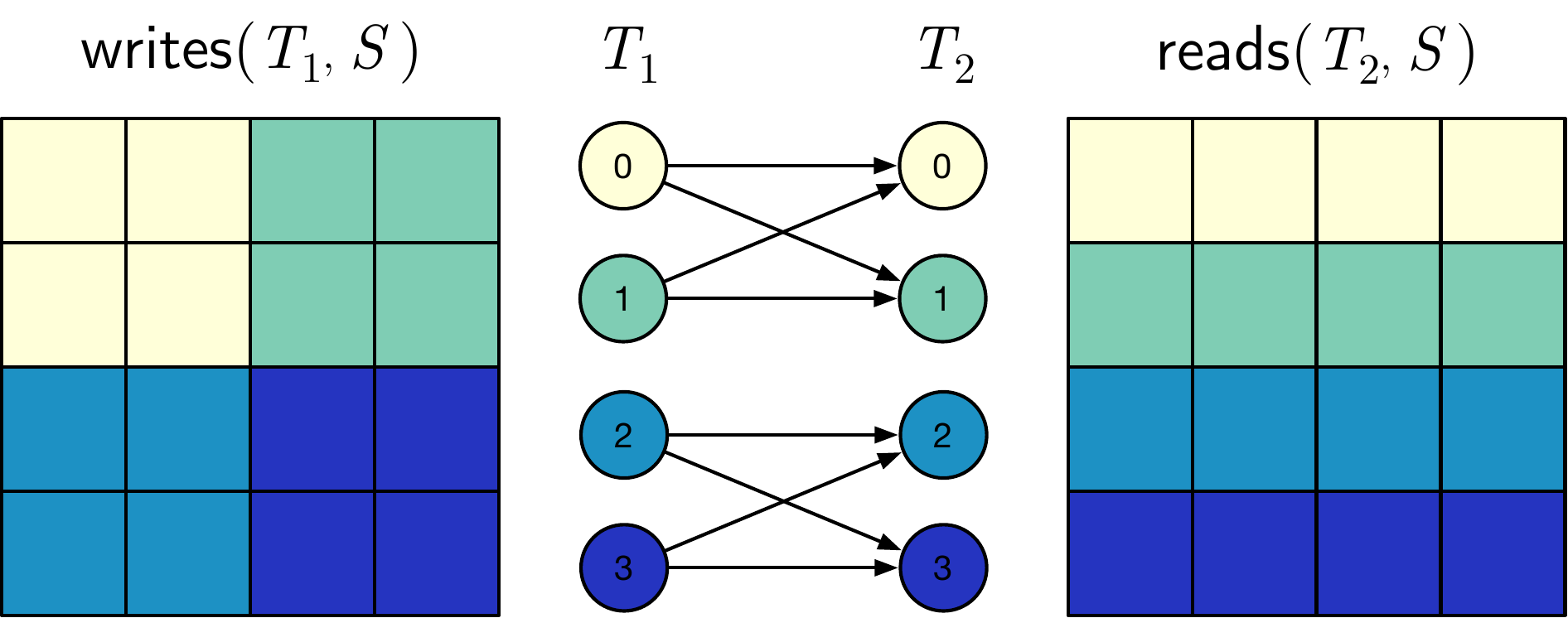}
    \caption{$\mathcal{D}(T_1, T_2)$ of writing to, then reading from different partitions.}
\end{subfigure}
\hfill
\caption{Visualization of dependence maps $\mathcal{D}(T_1, T_2)$.}
\label{fig:dep-map-viz}
\end{figure}

%

The dependencies between two index tasks $T_1$ and $T_2$ are defined by the pairwise dependencies
of their point tasks.
We capture these dependencies through a mapping between the points of $T_1$ and $T_2$ that represents
all of the point tasks in $T_2$ that depend on point tasks in $T_1$.
\Cref{fig:dep-map-viz} shows different dependence maps 
over the launch domain $(4,)$.
\begin{definition}
    For two index tasks $T_1$ and $T_2$, the \emph{dependence map} 
    $\mathcal{D}(T_1, T_2)$ is a function of type
    $\textsf{domain}(T_1) \rightarrow \mathcal{P}(\textsf{domain}(T_2))$, where
    $\forall p \in \textsf{domain}(T_1), \allowbreak \mathcal{D}(T_1, T_2)[p] = \{ p' \in \textsf{domain}(T_2) :
    \textsf{dep}(T^{p}_1, T^{p'}_2) \}$.
\end{definition}


    


Having characterized the dependencies between two distributed index tasks $T_1$ and $T_2$,
we can now define when fusion of $T_1$ and $T_2$ is valid.
$T_1$ and $T_2$ may be fused if the only dependencies that exist between their point tasks
are at most point-wise, as the processor that executes each point task does not need to communicate
with any other processors.

\begin{definition}\label{defn:fusion-possible}
    Index tasks $T_1$ and $T_2$ are fusible if 
    $\forall p, \mathcal{D}(T_1, T_2)[p]\allowbreak \subseteq \{p\}$.
\end{definition}

While \Cref{defn:fusion-possible} admits a simple dependency structure,
there are several subtleties in what tasks are fusible and the identification
of fusible tasks.
First, tasks with at most point-wise dependencies is a broader set than just tasks
that perform point-wise array operations.
Point-wise dependencies allow for simultaneous reads and writes of different stores
(\Cref{sec:motivating-example}) and multiple reductions to the same store.
While task dependencies may be at most point-wise, the computations within
the tasks are arbitrary computations that may be more complex than point-wise operations.
Next, identifying when at most point-wise dependencies exist between two index tasks
is non-trivial as tasks operate on arbitrarily aliasing distributed data.
We provide a framework to reason about fusion in this setting, allowing for fusion
to be performed between components within and across libraries.

\IGNORE{
\begin{itemize}
    \item Not just element-wise operations, at most point-wise dependencies -- important difference! Allows for aliasing reads, some writes mixed in, reduction fusion etc.
    \item Even this restriction is still an important class of optimizations, where fusion can bring large benefits to existing applications (point to evaluation).
    \item Detecting such point-wise cases is still difficult in the face of arbitrary aliasing, tasks coming from different parts of the same library, or different libraries all together.
\end{itemize}
}

\subsection{Fusion Algorithm}

A na\"ive algorithm for fusion might fully materialize $\mathcal{D}(T_1, T_2)$ to check
that the condition in \Cref{defn:fusion-possible} holds.
However, the computation required to materialize $\mathcal{D}(T_1, T_2)$ scales with the number of processors.
Even runtime systems like Legion do not materialize all of $\mathcal{D}$,
but instead leverage sophisticated algorithms to compute only the portion of $\mathcal{D}$
needed by each node~\cite{equivalence-sets}.
However, a key insight in our work is that to perform distributed task
fusion effectively, our analysis only needs to rule out cases where
$~\exists~p, \mathcal{D}(T_1, T_2)[p] \not\subseteq \{p\}$.
\name{}'s intermediate representation enables this analysis
to be performed in a scale-free manner.
%
%
Our algorithm for distributed task fusion identifies when
index tasks have point-wise dependencies through greedy application of
a set of \emph{fusion constraints} to identify a fusible prefix
of the task window.
We then build a fused task from the identified prefix.
%
%
We describe each of these components in turn, and then sketch a  
correctness proof in the next section.

\subsubsection{Fusion Constraints}

\name{} uses four constraints to identify when communication may
occur between distributed index tasks, i.e., when $\exists p,\allowbreak \mathcal{D}(T_1, T_2)[p] \allowbreak\not\subseteq \allowbreak\{p\}$.
The launch-domain-equivalence and true-dependence constraints have 
been described at a high level by prior work~\cite{shiv-fusion}.
We generalize these constraints from prior work, present formal definitions, 
and prove the correctness of our fusion algorithm.
\name{}'s fusion constraints are sound, but not complete---for example, leveraging
application knowledge could result in fusion opportunities that are out of scope
for \name{}.
\Cref{fig:fusion-constraint-definitions} presents each 
of the constraints used by \name{} by defining when a
provided sequence of tasks satisfy the constraint.

\textit{Launch Domain Equivalence.}
The first constraint checks that the tasks to be fused have the same launch domain.
%
%
Applications targeting \name{} may decompose their
computations across different launch domains, and 
data movement is generally required between different
decompositions.


\textit{True Dependence.}
The next constraint utilizes the partitions of stores and the privileges
with which they are accessed to identify communication
between index tasks caused by read-after-write dependencies.
%
%
The true-dependence constraint checks that if a task $T_i$ 
writes to a store $S$ through partition $P$, then it cannot be followed by a task $T_j$
that reads or writes to $S$ with an aliasing partition $P'$, as $T_j$ requires communication of the updated values
written by $T_i$.
However, operating on the same partition $P$ is permitted,
preserving point-wise dependencies between $T_i$ and $T_j$.
%

Our analysis relies on the ability to check whether two partitions
alias, which \name{} does through a constant-time equality
check between partitions.
Constant-time alias checking is possible
through the scale-free structure of \name{}'s IR and the 
syntactic grouping of partitions into structured kinds.
\name{} does not need to compute pairwise intersections of
the sub-stores accessed by the point tasks of considered index tasks,
a computation that scales quadratically with the number of processors.
Additionally, the alias analysis does not depend on the structure
of the partitions, as the constraints are defined without
knowing the syntactic kinds of each partition.
Finally, this aliasing check is not too coarse, since
partitions of different syntactic kinds nearly always
alias in practice.

\begin{figure}
\small
    \[
        \begin{array}{lll}

        \textsf{launch-domain-equivalence}([T_1, \ldots, T_n]) = & & \\
        \quad \forall i, \textsf{domain}(T_i) = \textsf{domain}(T_1) & &\\



        \textsf{true-dependence}([T_1, \ldots, T_n]) = &  &  \\
        \quad \forall~T_i~\text{s.t.}~\textsf{W}(T_i, (S, P)), && \\
        \quad \not\exists T_j~\text{s.t.}~\left(\textsf{R}(T_j, (S, P')) \vee \textsf{W}(T_j, (S, P'))\right)~\wedge i < j \wedge P \neq P' &&\\
        
%

        
        \textsf{anti-dependence}([T_1, \ldots, T_n]) = & & \\ 
        \quad \forall T_i ~\text{s.t.}~\textsf{R}(T_i, (S, P)), && \\
        \quad \not\exists T_j ~\text{s.t.}~\textsf{W}(T_j, (S, P'))~\wedge i < j \wedge P \neq P' &&\\



        \textsf{reduction}([T_1, \ldots, T_n]) = & &  \\
        \quad \forall T_i ~\text{s.t.}~\textsf{Rd}(T_i, (S, P)), && \\
        \quad \not\exists T_j ~\text{s.t.}~\left(\textsf{R}(T_j, (S, P')) \vee \textsf{W}(T_j, (S, P'))\right)~\wedge i \neq j &&\\


        \end{array}
    \]
          
    \caption{Fusion constraints employed by \name{} to identify potential communication between index tasks.}
    \label{fig:fusion-constraint-definitions}
\end{figure}



\textit{Anti-Dependence.}
The anti-dependence constraint ensures that $\mathcal{D}$ does not contain write-after-read
dependencies.
%
The constraint enforces that
if a task $T$ reads a store $S$, then any tasks that write to $S$
must write to the same distributed view as the read to be fused with $T$.
Thus, a fused task may read from multiple different distributed views
of a store (like the offset views of the stencil computation in \Cref{fig:cunumeric-stencil}),
but then cannot write to any of the views, as such an operation would require communication
of the written data.

\textit{Reduction.}
The reduction constraint makes sure that
viewing a partially reduced value is not allowed.
%
It does not permit a task that reads from or writes to a store to be
fused with a task performing a reduction to any view of the same store.

\subsubsection{Fused Task Construction}

Our fusion algorithm greedily applies the fusion constraints 
on the input task window to find its longest fusible prefix.
%
%
The true-dependence and anti-dependence constraints are verified through
a forwards dataflow analysis on the task window.
The analyses iterate through the candidate prefix of tasks, and track
the effects that each task applies to its argument stores.
Once a suitable prefix of the task window has been identified,
\name{} builds a fused task  that has all store arguments and
executes the same computation as the identified prefix of tasks.
%
%
The fused task's store arguments are constructed by reading all stores
read by tasks in the prefix, and similarly for the written to and reduced to stores.
Stores that are both read from and written to are promoted to the \textsf{Read-Write} privilege.
\name{} constructs the body of the fused task by composing the bodies
of each task in the prefix in program order---we further discuss this process
in \Cref{sec:kernel-fusion}.

\subsection{Proof of Correctness}


%
We now show that our algorithm correctly
fuses sequences of distributed index tasks.
We prove the following statement:

\begin{theorem}
    Given a window of tasks $[T_1, \ldots, T_n]$, our task fusion algorithm
    identifies a prefix $[T_1, \ldots, T_f]$ and produces a fused task $F$
    such that
    \begin{enumerate}
        \item $[T_1, \ldots, T_f]$ are fusible, and
        \item $F$ preserves all dependencies of the task sequence $[T_1, \ldots, T_f]$.
    \end{enumerate}
\end{theorem}

We provide a proof sketch for each component of the theorem.
To prove that $[T_1, \ldots, T_f]$ are fusible, we must show that
for each pair of tasks $T_i, T_j, i < j$ in $[T_1, \ldots, T_f]$,
$\forall~p,~\mathcal{D}(T_i, T_j)[p] \subseteq \{p\}$.
The launch-domain-equivalence constraint ensures that the dependence map is between
points of the same dimensionality.
For the sake of obtaining a contradiction, suppose
$\exists~p, p'$ such that $p \neq p'$ and $\textsf{depends}(T_i^p, T_j^{p'})$.
Then one of the three dependencies in \Cref{def:dependencies} must exist.
Suppose that the condition for true-dep is satisfied,
meaning that $\exists S, P, P'$ such that $S[P, p] \cap S[P', p']$
and $\textsf{W}(T_i, (S, P))$ and one of $\textsf{R}(T_j, (S, P'))$,
$\textsf{W}(T_j, (S, P'))$ or $\textsf{Rd}(T_j, (S, P')$ is true.
$\textsf{R}(T_j, (S, P'))$ or $\textsf{W}(T_j, (S, P'))$ are contradictions,
as the true-dependence constraint would disallow fusion.
$\textsf{Rd}(T_j, (S, P'))$ is a contradiction due to the reduction constraint.
Similar logic can be applied to other dependence cases.
Here, we show that our algorithm is sound by identifying cases where fusion is
possible---we do not claim completeness by proving the converse.

%

We have shown that all dependencies between index tasks are at most point-wise, 
so any $T_j^p$ can only depend on $T_i^p$, where $i < j$.
Since the fused task body is the composition of each task in $[T_1, \ldots, T_f]$ in program order,
all dependencies in $[T_1, \ldots, T_f]$ are preserved.

\subsection{Discussion}

Fusion at \name{}'s middle layer of abstraction is key
for a domain-agnostic analysis,
and for analysis scalability as the size of the machine increases.
We compare against fusion on high-level domain-specific libraries, 
and against fusion within lower-level runtime systems like Legion.

Domain-specific algorithms for fusion~\cite{jax, distal, spdistal, tensorflow, unity} are effective optimizations for individual distributed libraries.
%
Approaches that perform fusion on a set of domain-specific computations
use algorithms and analyses that are tied to the domain of computations being optimized,
especially analyses related to distributed memory.
As a result, these techniques do not readily generalize across libraries.
\name{} targets fusion in the more general case after computations have 
been decomposed into tasks in a domain-specific manner, enabling domain-agnostic analyses 
to find optimizations across  function and library boundaries.
We expect that domain-specific techniques may be used in conjunction with the analyses performed by \name{}.

While generality is lost when fusing operations within individual libraries,
scalability becomes a concern when analyzing lower-level program representations.
A key design decision in \name{}'s IR 
is that it is scale-free, as the representation of
parallel task groups and partitions of distributed data
are independent of the degree of parallelism.
This design enables \name{} to symbolically compute a conservative estimate of the aliasing relationships between distributed
data structures through constant-time queries, which are heavily used when defining
the fusion constraints in \Cref{fig:fusion-constraint-definitions}.
In contrast, lower-level systems like Legion represent partitions by explicitly mapping points to arbitrary sets of indices into the distributed data, scaling with the
number of pieces the data is partitioned into.
These representations are more flexible than \name{}'s, but
result in the aliasing relationship queries needed by a fusion algorithm to 
scale with the degree of available parallelism.

\section{Task Fusion Optimizations}\label{sec:fusion-optimizations}

%
Having described our algorithm for task fusion, we now describe optimizations
necessary for a practical implementation.
We show how to eliminate temporary distributed data structures (\Cref{sec:temp-elim})
and how to memoize the fusion analysis (\Cref{sec:memoization}).
Temporary elimination and memoization are widely applied optimizations;
we discuss how to perform these optimizations in a distributed, task-based setting.

\subsection{Temporary Store Elimination}\label{sec:temp-elim}

Once \name{} identifies a fusible prefix of tasks, 
stores that fusion has made temporary may be promoted into
task-local data.
Conversion of distributed data into task-local data is critical for realizing
the benefits of fusion, as task-local data can then be optimized
away (\Cref{sec:kernel-fusion}) to maximize reuse.
%
%
%
%

\begin{figure}
\begin{subfigure}[b]{0.49\linewidth}
\begin{center}
\begin{tabular}{c}
\begin{lstlisting}
import cupynumeric as np
x, y = np.zeros(n), np.ones(n)
flush_window()
z = 2.0 * x(*\label{line:task-start}*)
w = y + z
v = w ** 2(*\label{line:task-end}*)
norm = np.linalg.norm(
  w[len(w)//2:])
del x, y, z, w
flush_window()(*\label{line:tempelim-done}*)
\end{lstlisting}
\end{tabular}
\end{center}
\caption{\cunumeric{} code fragment.}
\label{fig:temp-code-cn}
\end{subfigure}
\hfill
\begin{subfigure}[b]{0.49\linewidth}
\begin{center}
\begin{tabular}{c}
\begin{lstlisting}
# Partitions and launch 
# domains excluded.
---
MULT([(x, R), (z, W)])
ADD([(y, R), (z, R, (w, W)])
POW([(w, R), (v, W)])
---
NORM([
 (w[len(w)//2:], R), (norm, Rd)
])
\end{lstlisting}
\end{tabular}
\end{center}
\caption{Emitted task stream.}
\label{fig:temp-code-stream}
\end{subfigure}
\caption{Example of distributed temporaries.}
\label{fig:temp-elim-exposition}
\end{figure}

To introduce when a store is temporary, consider 
the \cunumeric{} program in \Cref{fig:temp-code-cn} and the resulting task stream in \Cref{fig:temp-code-stream}.
This example introduces some new operations, specifically \texttt{flush\_window}, which
sends all pending tasks through \name{} to the underlying runtime system, and the
Python \texttt{del} operator, which drops references.
The program creates the stores \texttt{x}, \texttt{y}, \texttt{z}, \texttt{w}, and \texttt{v}.
Consider the program state after line~\ref{line:tempelim-done}:
the tasks that initialize \texttt{x} and \texttt{y} have 
executed, as the first \texttt{flush\_window} call sent those tasks to \name{}.
We note that there are no pending tasks outside the window, and future
tasks are ones the application may launch once the call to
\texttt{flush\_window} returns.
The fusion algorithm determines that the tasks issued by 
lines~\ref{line:task-start}--\ref{line:task-end} can be
fused, while the final \texttt{norm} must be excluded.
%
%
First, \texttt{v} is not temporary because the application holds a reference
to it, meaning that it could launch a task that reads \texttt{v} after the
call to \texttt{flush\_window()}.
%
Next, while the application has deleted its reference to \texttt{w}, the \texttt{norm} task reads
a piece of \texttt{w} and is still pending after the fused task, and thus must observe
any effects performed on \texttt{w}, meaning that \texttt{w} is not temporary.
The stores \texttt{x} and \texttt{y} are only read by the fused task, and thus are not temporary.
%
Only \texttt{z} is temporary because it is produced entirely within
the fused task and is not visible to the application or pending tasks.
%
We formalize this intuition as constraints that must be satisfied
for a store to be temporary.
\begin{definition}
    Given tasks $[T_1, \ldots, T_f, \ldots, T_n]$, a store $S$ is \emph{temporary} 
    in the fusion of $[T_1, \ldots, T_f]$ if
    \begin{enumerate}
        \item If $~\exists~T_j, P$ s.t. $\textsf{R}(T_j, (S, P))$, $\exists~T_i$ such that $i < j \wedge \textsf{W}(T_i, (S, P)) \wedge \textsf{covers}(S, P)$
        \item $\not\exists~T_k, P$ s.t. $k > f \wedge \textsf{R}(T_k, (S, P)) \vee \textsf{Rd}(T_k, (S, P))$
        \item $S$ has no live application references.
    \end{enumerate}
\end{definition}
The function $\textsf{covers}(S, P)$ is true when the partition $P$ contains all points
in the store $S$. 
The first two constraints check that the store's contents are entirely created
within the fused task and not used by any other existing task; these conditions are
checked through a forwards dataflow analysis of the task stream.
The third constraint ensures that the application can no longer view any effects on a store, checked
through a split reference counting scheme in the implementation of \name{}'s IR.
The split reference counting scheme separates references held by the application from references
held by \name{}'s runtime.
Temporary stores are demoted from a distributed allocation into 
a task-local allocation, as described in \Cref{sec:kernel-fusion}.

\subsection{Memoization of Analyses}\label{sec:memoization}

The final component of our distributed task fusion pipeline is memoization
analysis and code generation (\Cref{sec:kernel-fusion}).
The key challenge in memoization is allowing for the analyses to be replayed on
\emph{isomorphic} task streams rather than identical task streams.
%
Consider the streams of tasks in \Cref{fig:iso-non-iso-streams},
where partitions and launch domains are excluded. 
%

\begin{figure}
\begin{subfigure}[h]{1.05\linewidth}

\begin{subfigure}[h]{0.325\textwidth}
\begin{lstlisting}
T1([(S1,R), (S2,W)])
T2([(S2,R), (S1,W)])
T3([(S1,R), (S3,W)])
T4([(S3,R), (S1,W)])
\end{lstlisting}
\end{subfigure}
\hfill
\begin{subfigure}[h]{0.325\textwidth}
\begin{lstlisting}
T1([(S5,R), (S6,W)])
T2([(S6,R), (S5,W)])
T3([(S5,R), (S7,W)])
T4([(S7,R), (S5,W)])
\end{lstlisting}
\end{subfigure}
\hfill
\begin{subfigure}[h]{0.325\textwidth}
\begin{lstlisting}
T1([(S5,R), (S6,W)])
T2([(S6,R), (S5,W)])
T3([(~S7~,R), (S7,W)])
T4([(S7,R), (S5,W)])
\end{lstlisting}
\end{subfigure}
\caption{Two isomorphic task streams and one differing task stream.}
\label{fig:iso-non-iso-streams}
\end{subfigure}

\begin{subfigure}[h]{\linewidth}
\begin{subfigure}[h]{0.49\textwidth}
\begin{center}
\begin{tabular}{c}
\begin{lstlisting}
T1([(0,R), (1,W)])
T2([(1,R), (0,W)])
T3([(1,R), (2,W)])
T4([(2,R), (0,W)])
\end{lstlisting}
\end{tabular}
\end{center}
\end{subfigure}
\hfill
\begin{subfigure}[h]{0.49\textwidth}
\begin{center}
\begin{tabular}{c}
\begin{lstlisting}
T1([(0,R), (1,W)])
T2([(1,R), (0,W)])
T3([(~2~,R), (2,W)])
T4([(2,R), (0,W)])
\end{lstlisting}
\end{tabular}
\end{center}
\end{subfigure}
\caption{Canonical representations of isomorphic and differing streams.}
\label{fig:task-streams-memo}
\end{subfigure}

\caption{Example of task stream memoization.}

\end{figure}

\name{} may reuse the analysis results from the left stream in
\Cref{fig:iso-non-iso-streams} on the middle stream, as the pattern of stores among tasks is  isomorphic.
In contrast, the right task stream in \Cref{fig:iso-non-iso-streams} has a different pattern of stores
across tasks, particularly the use of \texttt{S7} in \texttt{T3}.
We observe that this problem is identical to \emph{alpha-equivalence}, where each
store argument is a bound variable.
We identify when two task streams are isomorphic within \name{} through
a conversion to and comparison on a canonical, De-Brujin index-like representation.
This representation is shown in \Cref{fig:task-streams-memo}.
%
%
A similar technique has previously been used to avoid enumerating instruction sequences equivalent up to register renaming~\cite{bansal-superopti}.

\section{Kernel Fusion}\label{sec:kernel-fusion}

\begin{figure*}
\begin{subfigure}[h]{0.24\textwidth}

\begin{subfigure}[h]{\textwidth}
\begin{lstlisting}[basicstyle=\fontsize{6}{8}\ttfamily]
func.func @kernel(
  %a: memref<?xf64>,
  %b: memref<?xf64>,
  %c: memref<?xf64>) {
  %dim = memref.dim %c, 0
  affine.for %i = 0 to %dim {
    %0 = affine.load %a[%i]
    %1 = affine.load %b[%i]
    %2 = arith.addf %0, %1
    affine.store %2, %c[%i] }}
\end{lstlisting}
\caption{MLIR generated for an element-wise addition.}   
\label{fig:elemwise-mlir-fragment}
\end{subfigure}
\end{subfigure}
\hfill
\begin{subfigure}[h]{0.24\textwidth}
\begin{lstlisting}[basicstyle=\fontsize{6}{8}\ttfamily]
func.func @fused_kernel(
  %a: memref<?xf64>,
  %b: memref<?xf64>,
  %c: memref<?xf64>,
  %d: memref<?xf64>,
  %e: memref<?xf64>) {
  %dim = memref.dim %e, 0
  affine.for %i = 0 to %dim {
    %0 = affine.load %a[%i]
    %1 = affine.load %b[%i]
    %2 = arith.addf %0, %1
    affine.store %2, %c[%i] }
  affine.for %i = 0 to %dim {
    %0 = affine.load %c[%i]
    %1 = affine.load %d[%i]
    %2 = arith.addf %0, %1
    affine.store %2, %e[%i] }}
\end{lstlisting}
\caption{Initial body of fused task.}
\label{fig:initial-fused-task}
\end{subfigure}
\hfill
\begin{subfigure}[h]{0.24\textwidth}
\begin{lstlisting}[basicstyle=\fontsize{6}{8}\ttfamily]
func.func @fused_kernel(
  %a: memref<?xf64>,
  %b: memref<?xf64>,
  %d: memref<?xf64>,
  %e: memref<?xf64>) {
  %dim = memref.dim %e, 0
  %c = memref.alloc %dim
  affine.for %i = 0 to %dim {
    %0 = affine.load %a[%i]
    %1 = affine.load %b[%i]
    %2 = arith.addf %0, %1
    affine.store %2, %c[%i] }
  affine.for %i = 0 to %dim {
    %0 = affine.load %c[%i]
    %1 = affine.load %d[%i]
    %2 = arith.addf %0, %1
    affine.store %2, %e[%i] }}
\end{lstlisting}
\caption{After temporary elimination.}
\label{fig:fused-task-temp-elim}
\end{subfigure}
\hfill
\begin{subfigure}[h]{0.24\textwidth}
\begin{lstlisting}[basicstyle=\fontsize{6}{8}\ttfamily]
func.func @fused_kernel(
  %a: memref<?xf64>,
  %b: memref<?xf64>,
  %d: memref<?xf64>,
  %e: memref<?xf64>) {
  %dim = memref.dim %e, 0
  affine.par %i = 0 to %dim {
    %0 = affine.load %a[%i]
    %1 = affine.load %b[%i]
    %2 = arith.addf %0, %1
    %3 = affine.load %d[%i]
    %4 = arith.addf %2, %3
    affine.store %2, %e[%i] }}
\end{lstlisting}
\caption{Fully optimized fused task.}
\label{fig:optimized-task}
\end{subfigure}
\caption{Fused MLIR kernel for three way element-wise addition traversing the compilation pipeline. The initial kernel is created by sequentially composing two of the generated task bodies in \Cref{fig:elemwise-mlir-fragment}.}
\end{figure*}

%
The final component of \name{} is a compilation stack to optimize fused tasks.
A high-level program representation is required to both
perform optimizations like loop fusion and to lower to
different backends like GPUs and multi-threaded CPUs.
%
We leverage the MLIR
compiler stack, which is extensible and is pre-packaged with
many common compiler analyses.
We first provide background on MLIR, and then
describe the code generation process and optimizations performed within \name{}.
We then discuss how \name{}'s architecture enables the separation
of reasoning about distributed programs from the optimization
of nested loops.

\subsection{MLIR Background}

We leverage MLIR~\cite{mlir} to build a JIT compiler for \name{}.
MLIR is an extension of LLVM~\cite{llvm} that
provides compiler infrastructure for program analyses on higher-level languages
than three-address code.
The most relevant component of this infrastructure to our work is the notion
of a \emph{dialect}, which is an intermediate representation that has user-defined
semantics.
A key aspect of dialects in MLIR is that a single MLIR program can contain types
and operations from multiple dialects, enabling the composition of dialects with
different semantics.
Compilers built using the MLIR framework run passes over programs that
either optimize the operations within a single dialect, or convert between dialects
to perform progressive lowering.
\name{}'s compiler leverages community-developed
dialects and passes to optimize and lower task bodies
into CPU and GPU code.

\subsection{Generator Functions}

To describe \name{}'s compiler, we walk through the stages that a fused task traverses.
\cunumeric{} and \legatesparse{} developers implement tasks by defining variants
that target CPUs or GPUs.
To use \name{}, developers register a \emph{generator} function
with \name{} that returns an MLIR fragment  describing
the task's computation.
%
We found the integration effort of adding these generator functions
to be modest, requiring 50--100 lines of C++ code per operation.
We emphasize that only library developers, not end users, must develop
MLIR kernels for tasks.
Additionally, the integration effort was incremental---as more tasks
were implemented with MLIR generators, \name{} could exploit more fusion.
An example generated MLIR fragment by \cunumeric{} for
an element-wise addition operation is shown in \Cref{fig:elemwise-mlir-fragment}.

The generated MLIR fragment in \Cref{fig:elemwise-mlir-fragment} contains multiple
dialects: 1) stores are mapped onto the \texttt{memref} dialect, which provides
stronger aliasing guarantees than raw pointers; 2) dense iteration
is mapped onto the \texttt{affine} dialect, a target for 
polyhedral compilation~\cite{mlir-affine-polyhedral}; and 3) the computation itself is
mapped onto the \texttt{arith} dialect, containing arithmetic operations.
Using MLIR, other dialects can be used to express higher level operations,
like dense and sparse tensor algebra with the \texttt{linalg} and \texttt{sparse\_tensor}
dialects.




\subsection{Compilation Pipeline}

When \name{} identifies that a sequence of tasks may be fused, it invokes
each task generator and constructs an MLIR module
containing the body of each task in the original program order.
\Cref{fig:initial-fused-task} shows a fused task for the \cunumeric{} computation 
\texttt{c = a + b; e = c + d}, where all variables represent distributed vectors.
This program originally has two index tasks (one for each add operation)
which are fused into a single index task where the original task bodies 
(the MLIR in \Cref{fig:elemwise-mlir-fragment}) appear sequentially in the fused task.
Before optimization of the task body, \name{} first 
promotes distributed data into task-local allocations, resulting in
\Cref{fig:fused-task-temp-elim}.

After elimination of temporary stores, we apply passes that fuse and parallelize nested loops, and remove
task-local temporary allocations to yield the optimized code in \Cref{fig:optimized-task}.
The generated kernel is the ideal implementation for the original
program: the separate loops of the
original task bodies have been combined into a single pass
over the vectors, and the temporary \texttt{c} has been eliminated.
The optimized kernel is then lowered to GPU kernel launches or
OpenMP parallel regions.

In this work, we leverage polyhedral optimizations~\cite{mlir-affine-polyhedral, uday-thesis} to perform fusion and parallelization of loops in kernels.
However, with higher level dialects in MLIR, various domain-specific kernel fusion
techniques (see \Cref{sec:related-work}) could be leveraged within a fused task body.
We consider the exact kernel fusion techniques used to be orthogonal to our work.

\subsection{Qualitative Benefits}

%
We note several qualitative benefits of our system architecture
in contrast to approaches that attempt to optimize distributed programs
entirely through analysis of imperative code.
A key design decision of \name{} is to leverage a distributed data model 
in a scale-free IR
of computation that enables cheap dependence analysis between distributed computations.
Separating out the reasoning about distributed computation
avoids intertwining loop optimizations with distributed communication analyses,
allowing the loop optimizations to remain unaware of the distributed context.
This separation also allows for information gained during the distributed analysis
phase to be used in code generation: properties such as array non-aliasing
are provided to the MLIR optimization passes to generate better code.
Finally, the separation of distributed computation into tasks means
that \name{} does not need to
identify optimizable fragments of static source code.

\section{Evaluation}

\textit{Experimental Setup.}
We evaluate the performance of \name{} on a cluster of NVIDIA A100 DGX SuperPOD nodes.
Each node has 8 A100 GPUs with 80GB of memory, connected by NVLink and NVSwitch 
connections, and a dual socket, 128 core AMD 7742 Rome CPU with 2TB of memory.
Each node is connected via an InfiniBand connection through 8 NICs.

For each experiment, we perform a weak-scaling study, 
and report the throughput achieved per processor.
A weak-scaling study increases the problem size as the size
of the target machine grows to keep the problem size per
processor constant.
Each reported value is the result of performing 12 runs, dropping the
fastest and slowest runs, and then computing the average of the remaining 10 runs.
In weak-scaling experiment (\Cref{sec:weak-scaling}), we 
exclude a set of warmup iterations from timing to measure the steady-state
performance with and without \name{}.
We separately evaluate the overhead that \name{} imposes due to compilation
in \Cref{sec:compilation-time}.

\textit{Overview.}
We evaluate \name{} on unmodified, open source \cunumeric{} and \legatesparse{}
applications, from microbenchmarks to full applications.
Many of these applications have appeared in prior publications~\cite{cunumeric, legate-sparse},
and range from tens to thousands of lines of Python.
The unique capabilities of \cunumeric{} and \legatesparse{}
enable these pure Python applications with dynamic and
data-dependent behavior to be scaled across
multiple nodes of multiple GPUs.
An overview of each application is in \Cref{fig:exp-metadata}.
%
%
We compare each application's performance when run
with and without \name{} --- no changes to the application are needed to enable \name{}.
For some applications, a suitable baseline written in the industry-standard
PETSc~\cite{petsc} library for distributed sparse linear algebra already exists,
and we compare against those baselines.
For other applications, we compare against manually optimized implementations
by the original authors.
However, some full \cunumeric{} applications have no baseline other than 
when run without name---these applications are sufficiently complex that
developing an independent high-performance distributed, multi-GPU 
implementation is not feasible.
We show that when fusion opportunities are available, \name{} can exploit
them to find speedups in unmodified, distributed applications.
\name{} enables high-level programs to equal, and in many cases improve on,
the performance of hand-optimized code.

%

We do not ablate on the optimizations
in \Cref{sec:fusion-optimizations}, as temporary elimination is
essential for speedup with kernel fusion and
memoization is a requirement for a practical implementation.
We do not compare against the work of
Sundram et al.~\cite{shiv-fusion}, which performs only task-fusion, as the
version of \cunumeric{} they used is older and would not 
be a fair comparison.
However, we have evaluated \name{} with only task fusion and found
that it did not yield speedups on our benchmarks.
Task fusion alone can only reduce runtime overhead, and the
task granularity of our benchmarks is larger than the
minimum effective task granularity~\cite{task-bench} of Legion (1ms per task).
The window sizes shown in \Cref{fig:exp-metadata} were selected
automatically by \name{} through a process that
increases the window size when all tasks in the current window size
were fused.
As a result, these window sizes enable the maximum amount of fusion
possible in each application.
Finally, our benchmarks issue index tasks
that have one point per GPU, so computations are not
over-decomposed.

\begin{figure}
\scriptsize

\centering
\begin{tabular}{|c|c|c|c|c|}
\hline

\textbf{Benchmark} & \thead{Tasks per \\ Iteration} & \thead{Tasks per \\ Iteration (Fused)} & \thead{Avg Task \\ Length (ms)} & \thead{Window\\Size} \\
\hline

Black-Scholes & 67 & 1 & 5.3 & 70 \\
\hline
Jacobi & 3 & 2 & 5.3 & 5 \\
\hline
CG & 12.1 & 4.1 & 1.9 & 10 \\
\hline
BiCGSTAB & 27.1 & 8.1 & 1.7 & 15 \\
\hline
GMG & 24.1 & 11.1 & 1.8 & 15 \\
\hline
CFD & 378 & 40.7 & 1.1 & 30 \\
\hline
TorchSWE & 276.5 & 152.8 & 1.4 & 30 \\
\hline
\end{tabular}
\caption{Index tasks per iteration with and without fusion. Task count is not always whole as iterations may launch different tasks, or fusion occurs across iteration boundaries. Reported task granularities are from unfused single-GPU executions. Window size was selected by Diffuse.}
\label{fig:exp-metadata}
\end{figure}

\subsection{Weak Scaling Experiments}\label{sec:weak-scaling}

\begin{figure}
\begin{subfigure}{\linewidth}
\centering
\includegraphics[width=0.75\textwidth]{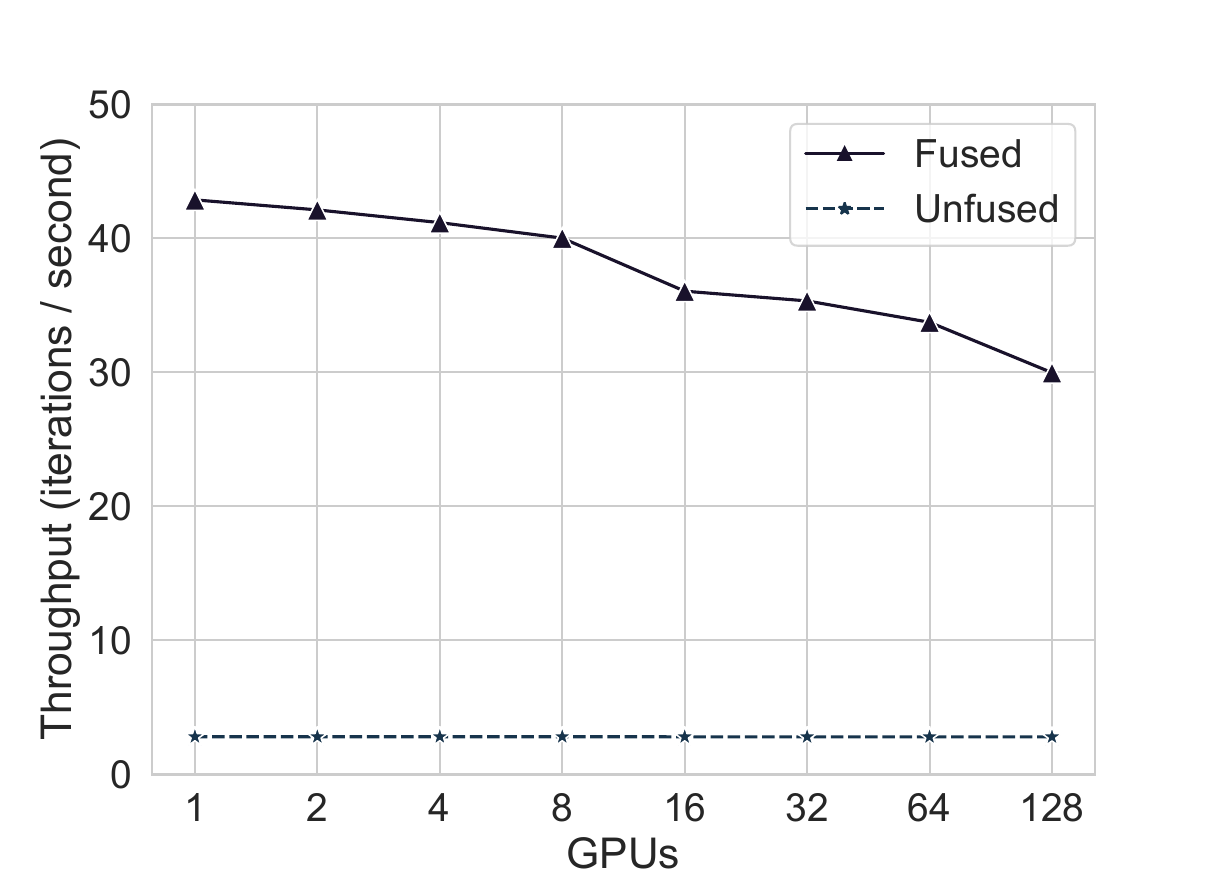}
\caption{Black-Scholes}
\label{fig:black-scholes}
\end{subfigure}
\begin{subfigure}{\linewidth}
\centering
\includegraphics[width=0.75\textwidth]{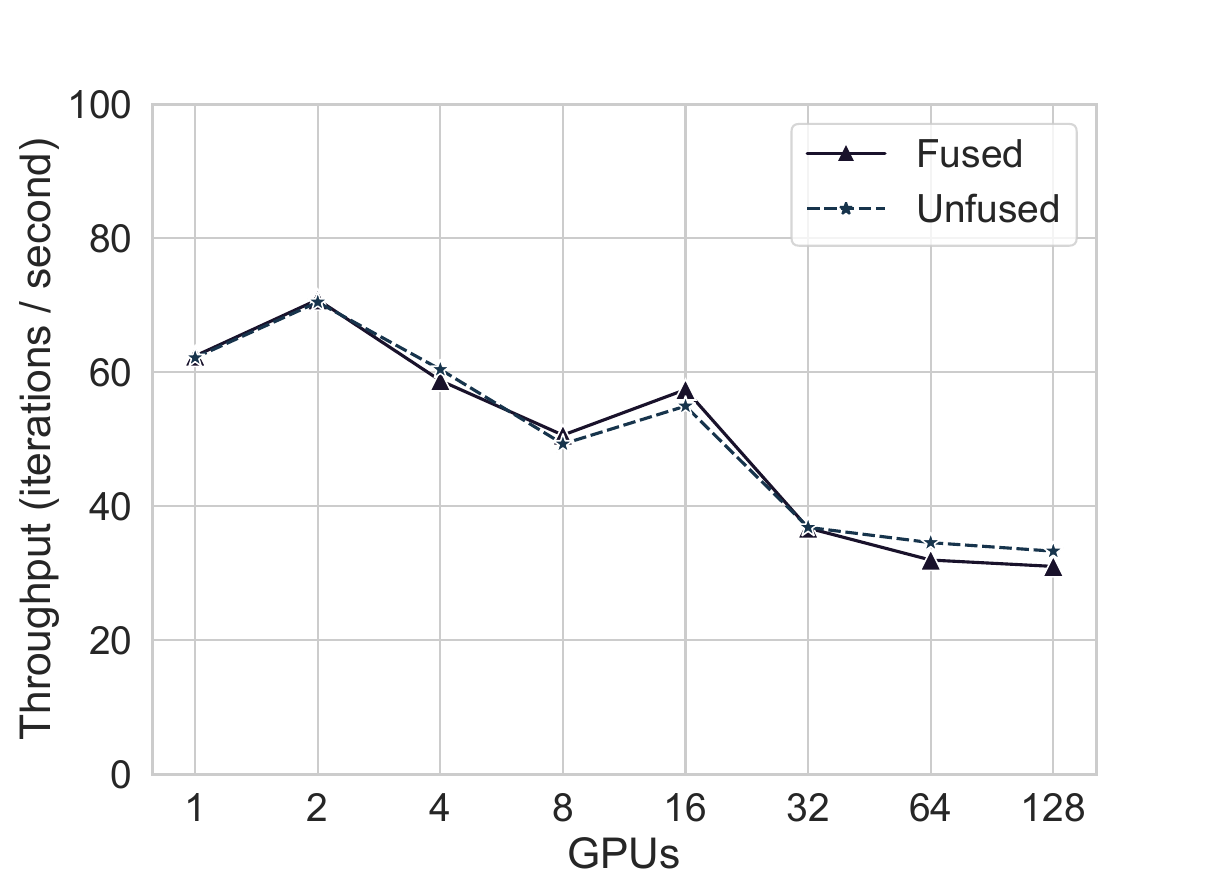}
\caption{Jacobi Iteration}
\label{fig:jacobi}
\end{subfigure}
\caption{Microbenchmark weak scaling (higher is better).}
\end{figure}

\textit{Black-Scholes.}
The Black-Scholes option pricing benchmark is a trivially-parallel micro-benchmark
that contains a sequence of 67 data-parallel, and thus fusible, operations.
It is a micro-benchmark that provides a reference point on potential improvement
when the entire application is amenable to fusion.
%
%
\Cref{fig:black-scholes} shows that \name{} achieves a 
10.7x speedup over the unfused program on 128 GPUs,
as the fused program is a single task containing a single GPU kernel
making one pass over the data, greatly increasing the arithmetic 
intensity of the computation.

\textit{Dense Jacobi Iteration.}
Unlike Black-Scholes, dense Jacobi iteration has negligible
potential benefit from fusion.
Jacobi iteration consists of a dense
matrix-vector multiplication and
two fusible vector operations that are negligible in runtime.
This benchmark shows that our analyses do not
have a significant negative impact on performance when there is no fusion.
\name{} achieves 0.93--1.08x of the performance
of the unfused Jacobi iteration in \Cref{fig:jacobi}, where we believe
the slight improvement is due to experimental variability.

\begin{figure}
\begin{subfigure}{\linewidth}
\centering
\includegraphics[width=0.75\textwidth]{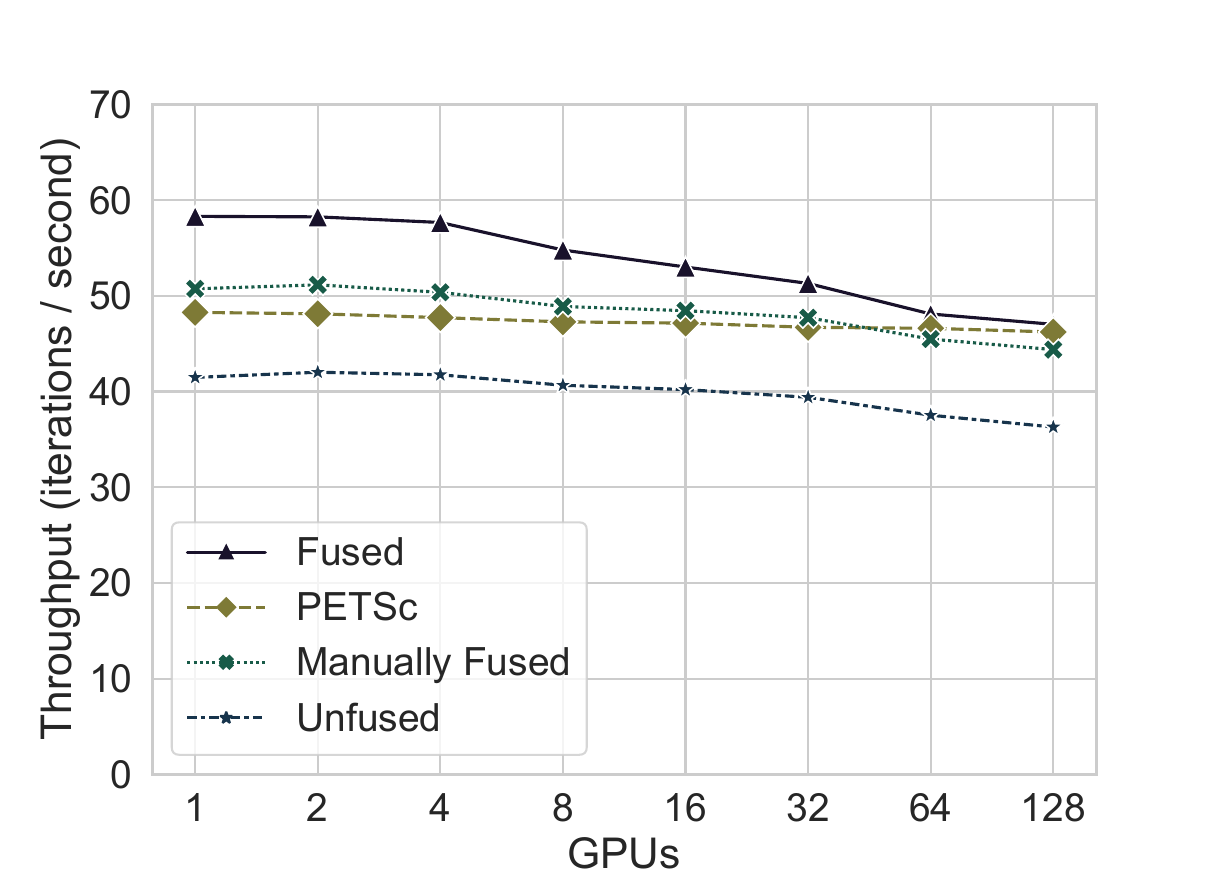}
\caption{CG}
\label{fig:cg}
\end{subfigure}
\begin{subfigure}{\linewidth}
\centering
\includegraphics[width=0.75\textwidth]{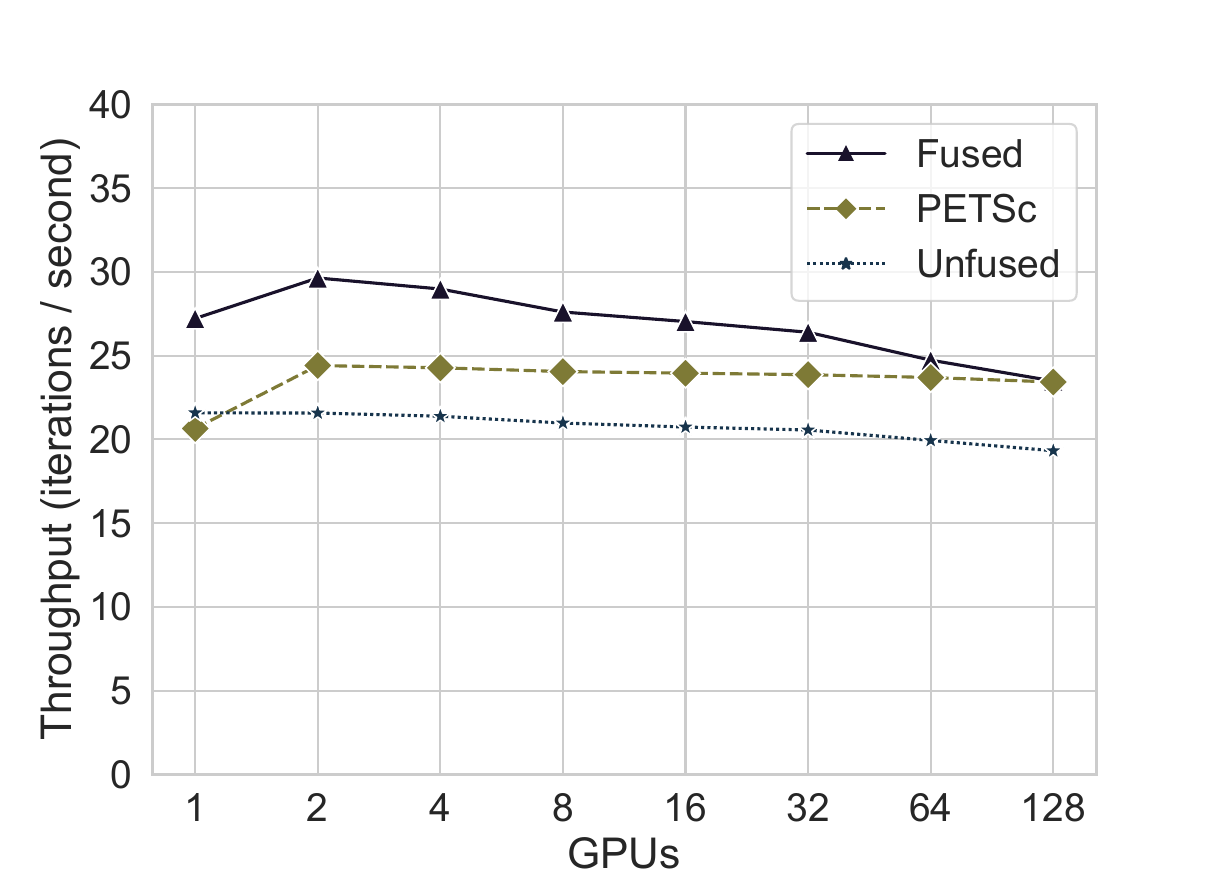}
\caption{BiCGSTAB}
\label{fig:bicgstab}
\end{subfigure}
\caption{Weak scaling of linear solvers (higher is better).}
\end{figure}

\begin{figure*}
\hfill
\begin{subfigure}[h]{0.33\linewidth}
\centering
\includegraphics[width=\textwidth]{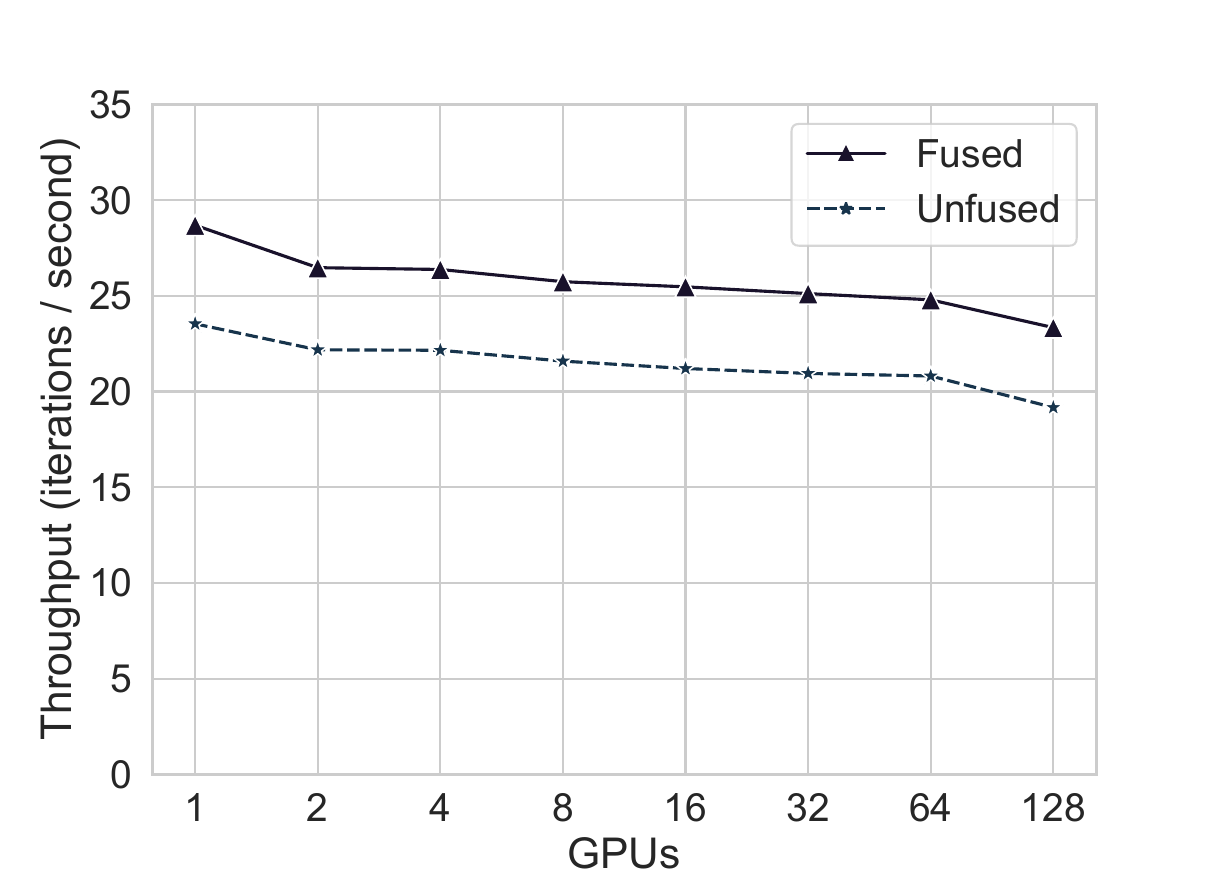}
\caption{GMG}
\label{fig:gmg}
\end{subfigure}\hfill
\begin{subfigure}[h]{0.33\linewidth}
\centering
\includegraphics[width=\textwidth]{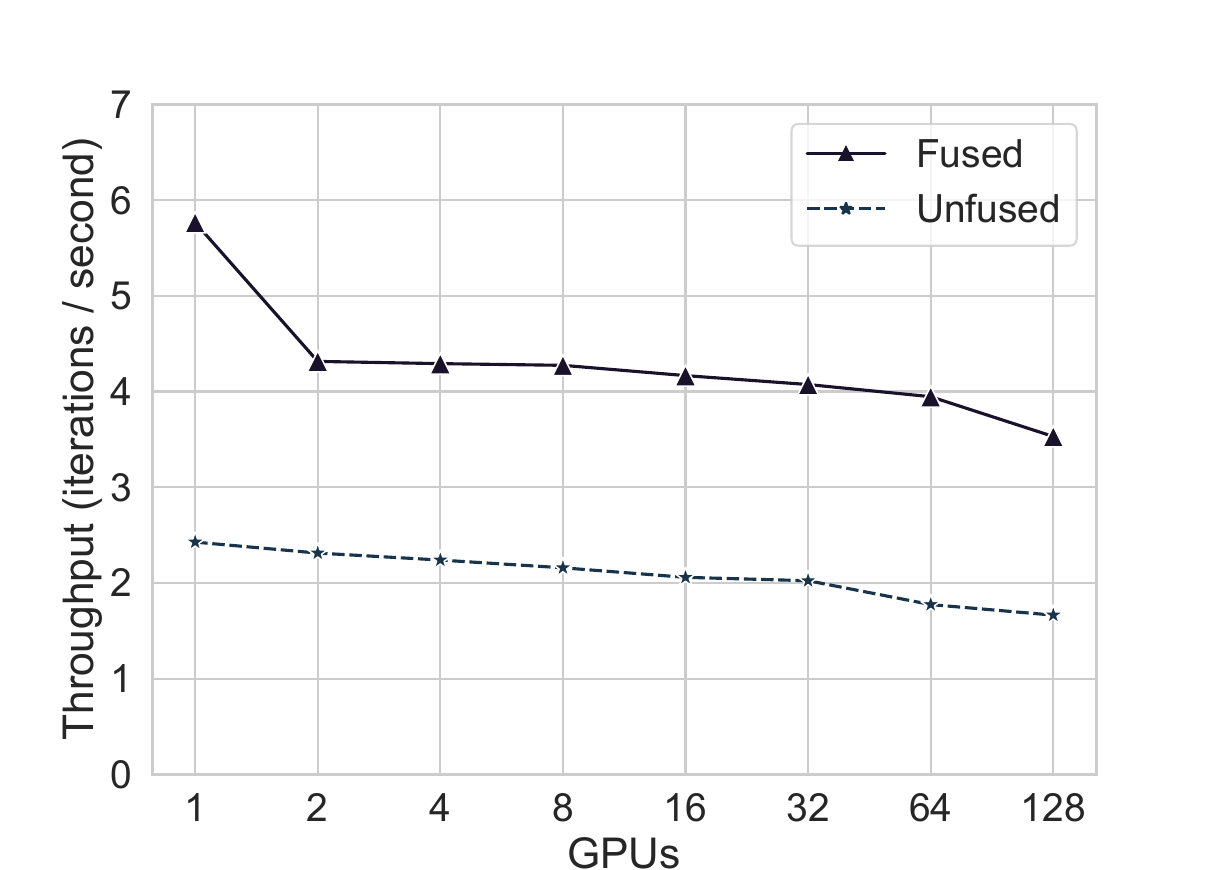}
\caption{CFD}
\label{fig:cfd}
\end{subfigure}\hfill
\begin{subfigure}[h]{0.33\linewidth}
\centering
\includegraphics[width=\textwidth]{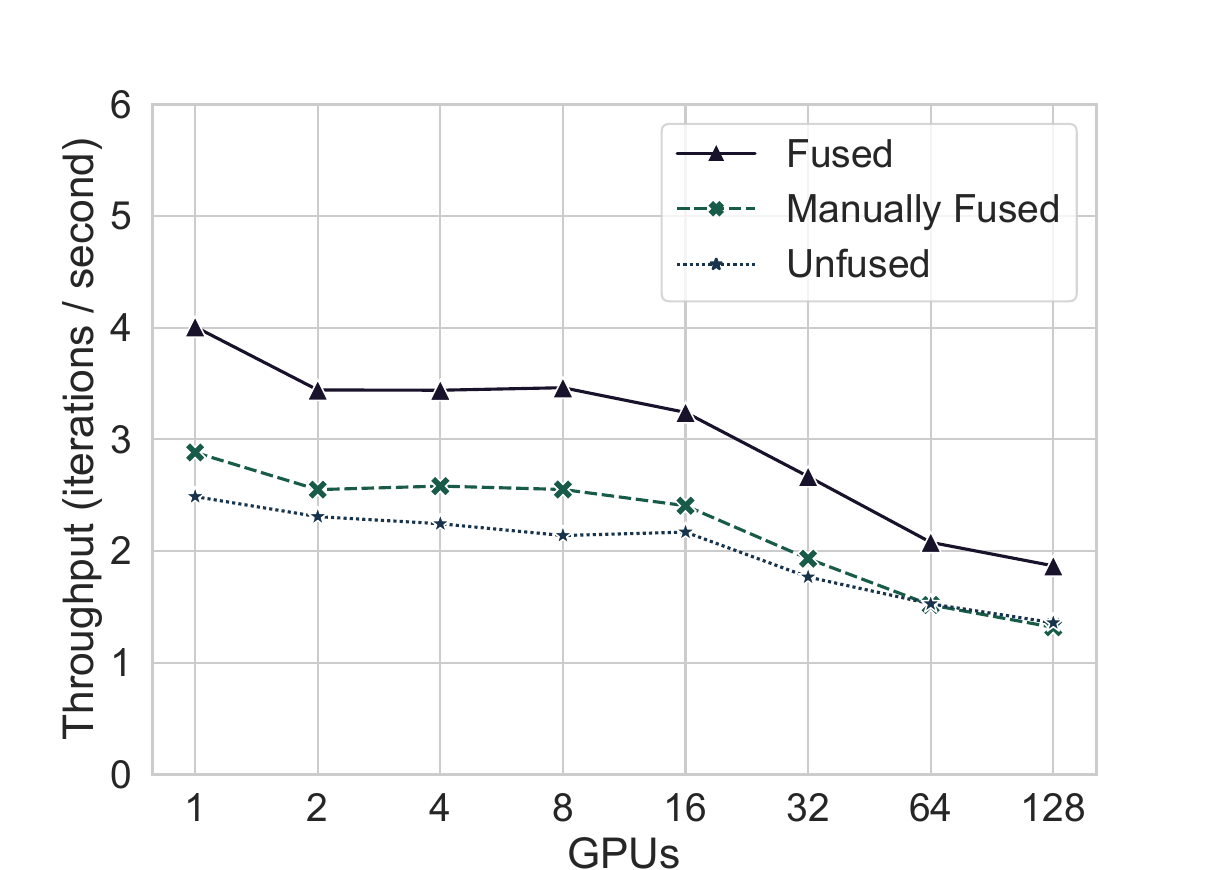}
\caption{TorchSWE}
\label{fig:torchswe}
\end{subfigure}\hfill
\hfill

\caption{Weak scaling of full applications (higher is better).}
\end{figure*}

\textit{Sparse Krylov Solvers.}
We evaluate sparse Krylov solvers implemented with
\cunumeric{} and \legatesparse{}, namely Conjugate Gradient (CG)
and Bi-Conjugate Gradient Stabilized (BiCGSTAB).
%
%
The PETSc benchmark implementations are implemented in MPI+C using
PETSc's API.
%
%
To perform a controlled comparison against PETSc, we modify 
\legatesparse{} to perform a similar optimization as PETSc, where the non-zero coordinates
in each sparse matrix partition are stored as 32-bit integers instead of 64-bit integers.\footnote{PETSc stores coordinates in 32-bit integers even when 64-bit integers
are requested at build time, affecting the performance of the SpMV kernel.}

The original implementation of CG in \legatesparse{} had been optimized manually
to perform many of the optimizations that \name{} does automatically.
As a result, the implementation no longer 
resembled the high-level description of CG.
We compare against this manually fused implementation,
a naturally written implementation using \cunumeric{}
and \legatesparse{}, and PETSc.
\Cref{fig:cg} shows that \name{} automatically optimizes the naturally
written CG so that it runs faster than both the manually
optimized version and PETSc.
\name{} finds additional
fusion opportunities by fusing
AXPY's and dot-products from different iterations.

We implement an unfused version of BiCGSTAB in
\cunumeric{} and \legatesparse{} and compare against PETSc.
%
\Cref{fig:bicgstab} shows that \name{} accelerates the high-level
implementation of BiCGSTAB to outperform the unfused version by
1.31x on average (geo-mean) and PETSc by 1.15x on average (geo-mean).
%
PETSc exposes several fused kernels to users for use in
building iterative solvers, but these kernels can quickly become
complicated and esoteric\footnote{Such as \texttt{VecAXPBYPCZ} in BiCGSTAB (\url{https://petsc.org/main/manualpages/Vec/VecAXPBYPCZ/}).}.
In contrast, \name{} enables users to write high-level programs in \cunumeric{}
and \legatesparse{} and then derives optimized kernels
for efficient execution.

\textit{Geometric Multi-Grid Solver (GMG).}
Moving from smaller benchmarks to full applications, we apply \name{} to
a Geometric Multi-Grid (GMG) solver developed in \legatesparse{}.
The GMG solver is a CG-based iterative solver with a V-cycling
preconditioner, the injection restriction operator, and a weighted
Jacobi smoother.
As with the previous benchmarks, using \name{} with the more complex solver
required no changes to user-facing code, and results in a 1.2x speedup
over the original implementation, as seen in \Cref{fig:gmg}.

\textit{Computational Fluid Dynamics (CFD).}
%
%
We apply \name{} to a \cunumeric{} application that solves the Navier-Stokes
equations for 2D channel flow~\cite{cfdemo}.
The application performs element-wise operations on aliasing
slices of distributed arrays, exposing opportunities for fusion.
\name{} finds between 1.8x--2.3x speedup over the original 
implementation, as shown in \Cref{fig:cfd}.
\name{} achieves higher speedup on a single GPU
than on multiple GPUs.
On a single GPU, data is not partitioned,
enabling longer sequences of tasks to satisfy fusion constraints.
On multiple GPUs, the dependencies caused by
aliasing data reduce the opportunities for fusion.



\textit{Shallow Water Equation Solver (TorchSWE).}
Our final benchmark application is also our most complex: the \cunumeric{} port 
of the TorchSWE shallow-water equation solver~\cite{torchswe}.
We compare against the original \cunumeric{} port, as well as a version
that the \cunumeric{} developers manually optimized using \texttt{numpy.vectorize}.
The \texttt{vectorize} utility JIT-compiles a user-defined
element-wise operator, doing some
of the optimizations that \name{} performs automatically.
\Cref{fig:torchswe} shows the performance of TorchSWE with \name{} compared to these baselines.
\name{} achieves a 1.61x speedup on average (geo-mean) over the unfused TorchSWE, and a 1.35x speedup on average (geo-mean) over the manually vectorized version (labeled with ``Manually Fused'' in \Cref{fig:torchswe}).
Since \name{} is analyzing the entire application,
it can find fusion opportunities missed by developers optimizing the program by hand.

\subsection{Compilation Time}\label{sec:compilation-time}

\begin{figure}

\scriptsize

\centering

\begin{tabular}{|c|c|c|c|}

\hline
\textbf{Benchmark} & \textbf{Standard (s)} & \textbf{Compiled (s)} & \textbf{Breakeven Iterations} \\
\hline

Black-Scholes & 0.38 & 0.06 & N/A \\
\hline
Jacobi & 0.53 & 0.43 & N/A \\
\hline
CG & 0.67 & 1.30 & 99.44 \\
\hline
BiCGSTAB & 1.26 & 2.19 & 80.43 \\
\hline
GMG & 0.49 & 1.38 & 118.75 \\
\hline
CFD & 5.10 & 10.89 & 25.21 \\
\hline
TorchSWE & 0.97 & 8.82 & 43.88 \\
\hline
\end{tabular}
\caption{Warmup times on 8 GPUs.}
\label{fig:comptime}
\end{figure}

We measure the overhead that
\name{}'s compilation imposes on overall runtime.
When evaluating our benchmarks, we compute the throughput
after warmup iterations have concluded.
%
To measure the effect of compilation, we measure the warmup time with and
without compilation, using the window sizes reported in \Cref{fig:exp-metadata}.
We then compute the number of iterations required for the fused version
to be faster than the unfused version of the application when
including the warmup compilation time.
The results are shown in \Cref{fig:comptime}; 
\name{}'s compilation times are modest, requiring 25--119 iterations
to amortize the cost of compilation.
The fused Black-Scholes computation is so much faster than the unfused version
that a single iteration is sufficient to amortize compilation.
For Jacobi, compilation time was overlapped with expensive dense matrix-vector
multiply kernel, and thus not exposed in the warmup.
As seen in \Cref{fig:jacobi}, due to experimental variation, the fused and unfused
versions of Jacobi are slightly faster or slower than each other on different GPU counts.
These costs are especially reasonable as scientific applications like the ones we evaluated
would be run in production for thousands to millions of iterations.
In the future, a production-grade implementation of \name{} could maintain
a cache of compiled kernels on disk, rather than in memory, and pay the compilation
cost only the first time the application is run.

\section{Related Work}\label{sec:related-work}


\textit{Task Fusion.}
Task fusion is a widely applied technique in parallel computing to
reduce the overheads of parallelism~\cite{dask-fusion, mapreduce-fusion, noll-gross, zhao-sarkar,rosa-binder,sam-parallel-management}.
%
%
Most prior work considers the fusion of individual tasks---in this work, we 
consider a more complex variant of task fusion, the fusion
of groups of distributed tasks, which is challenging due to the dependencies
that exist between distributed tasks.
The most related work is that of Sundram et al.~\cite{shiv-fusion},
which identifies the problem and provides an initial solution for detecting when
fusion of index tasks is possible.
We improve on this work by developing a formal model for reasoning about
distributed tasks, identifying new constraints
on fusion, and proving that the set is sufficient.
We then pair task fusion with a JIT compiler 
to fuse the task bodies, enabling \name{} to achieve significantly larger speedups than
just task fusion, as more potential benefits than runtime overhead removal are possible.

\textit{Kernel Fusion.}
%
%
Nested loop fusion in imperative, array-based programs is well-studied~\cite{allen-cocke-optimizations, kennedy-fusion, darte-fusion, udayb-thesis}.
Our work combines loop fusion with the data and computational models of a tasking
runtime to enable kernel fusion in a distributed environment.
Kernel fusion has also been explored heavily in different domains.
Deforestation approaches aim to remove temporary lists and 
trees in functional programs~\cite{wadler-deforestation}.
Fusion in collection-oriented languages
combines operations like map and reduce into single passes
over data structures~\cite{sam-sequences, guy-size-inference, parallel-patterns, monad-comprehensions, gill-deforestation}.
Various compilers have been developed
to generate fused code for operations over dense~\cite{halide, tvm, tensor-comprehensions} and sparse tensors~\cite{taco,bik2022compiler}.
Machine learning frameworks perform operator
fusion within neural networks~\cite{taso, dnnfusion, xla, jax, hfuse}.
%
%
Our work provides a domain-agnostic framework for identifying fusion
in streams of distributed tasks, and could leverage these techniques for kernel fusion.

\textit{Efficient Composition of Parallel Software.}
\name{} aims to efficiently compose operations
within and across distributed libraries.
Some recent projects have tackled the problem of efficient composition;
we discuss each in turn.
Weld~\cite{weld} provides a loop-based IR in which users can define
single-node library computations, and a runtime system that optimizes the IR to enable
cross-function and cross-library optimizations.
Split Annotations~\cite{split-annotations} provides partitioning
annotations for users to attach to library functions, and
uses these annotations to run cache-sized batches of the functions
to maximize data reuse.
%
Both Weld and Split Annotations target a similar problem as \name{}, but would require a model of distributed data like the one we propose to safely perform optimizations in a distributed setting.
%
DaCe~\cite{dace} is a compiler that leverages an IR called
Stateful Dataflow MultiGraphs to perform optimizations like fusion on Python/NumPy programs.
Distributed programs in DaCe are explicitly parallel, including manual communication
with libraries like MPI, which requires different kinds of analyses.

Jax~\cite{jax} and PyTorch~\cite{pytorch-compiler} are machine-learning systems that
compile \numpy{}-like descriptions of neural networks to perform optimizations like
fusion and automatic differentiation.
Systems like Jax and PyTorch accept structured program representations (neural network graphs)
and apply optimizations that leverage domain-specific knowledge, many of which are not
possible for \name{} to perform.
In contrast, \name{} only leverages the privilege information about tasks to perform
optimizations, and allows for description of programs with complex aliasing and mutation
that are not possible to represent in ML systems, like the CFD or TorchSWE simulations.
We consider \name{} to be a different point in the design space than these ML systems,
focusing on fusion in a more general setting without application-specific knowledge.

\textit{Distributed Runtime Systems.}
\name{} uses a scale-free IR
to efficiently perform distributed dependence and alias analyses.
This is similar to Index Launches~\cite{index-launches},
a representation of distributed tasks that compresses the degree of parallelism.
\name{}'s model of distributed data supports \emph{content-based coherence}, meaning
that the same data may be referred to in multiple different ways.
Legion~\cite{legion}, which \name{} builds upon, is a
system that supports content-based coherence of distributed data.
Legion exposes a more general interface for partitioning data, allowing
a partition to contain arbitrary subsets.
Legion then uses sophisticated algorithms for
computing dependencies between tasks and maintaining coherence of distributed data~\cite{equivalence-sets}.
Legion's flexible data model and support for precise dependence analysis at scale
are critical features for building libraries like \cunumeric{} and Legate Sparse.
Supporting Legion's flexible data model is a key challenge in \name{}, as
libraries that target Legion depend on this capability.
\name{}'s restricted data representation and goal of only fusion enable
compact analyses for the dependence and coherence problems.
In systems without content-based coherence, simpler approaches than ours may suffice,
as aliasing distributed data is no longer a concern.

\IGNORE{
\paragraph{Just-In-Time Compilation}

\TODO{JITs -- dynamioRIO, java JIT (self project), julia JIT, jits in databases?}
}

\IGNORE{

\section{Future Work}

\begin{enumerate}
    \item Multiple MLIR dialects in the same program for cross-library optimizations.
    \item Adjusting other stages of the pipeline to make more opportunities for fusion, such as improving the auto-partitioner to partition in batches to choose something that minimizes communication, and thus increases the fusion opporutinities.
\end{enumerate}

}

\section{Conclusion}

We introduced \name{}, a system that performs task and kernel fusion on streams of
distributed tasks, enabling optimizations that improve data reuse and remove
allocations of distributed data structures in end user programs.
\name{} leverages a scale-free intermediate representation of distributed computation
and data to perform these analyses in a scalable manner.
These techniques enable \name{} to compose computations in and across  \cunumeric{} and \legatesparse{}, matching or
exceeding the performance of hand-tuned code.
\IGNORE{
We believe the techniques developed in \name{} are the first steps towards a full suite
of optimizations to efficiently compose distributed and accelerated software.
}

\section*{Acknowledgements}

We thank Scott Kovach for his assistance in formalizing the fusion correctness proof.
We thank Shriram Jagannathan and Irina Demeshko for their assistance in running
the vectorized TorchSWE benchmark.
We thank (in no particular order) David Broman, James Dong, AJ Root, Scott Kovach, Parthiv Krishna, Benjamin Driscoll, Olivia Hsu, Marco Siracusa, Rubens Lacouture for their comments and
discussions on early stages of this manuscript.
Rohan Yadav was supported by an NVIDIA Graduate Fellowship, and part of this work was done while
Rohan Yadav was an intern at NVIDIA Research.
This work was in part supported by the National Science Foundation under Grant CCF-2216964.

\bibliographystyle{ACM-Reference-Format}
\balance
\bibliography{main}

\end{document}